\documentclass[12pt]{article}
\usepackage{epsf}
\usepackage{graphicx}
\usepackage{lscape}
\usepackage{amssymb}
\usepackage{pazh}
\usepackage{natbib}

\tightenlines

\begin{document}

{\footnotesize Astronomy Letters, Vol. 33, No. 7, 2007, pp. 437-454.
Translated from Pis'ma v Astronomicheskii Zhurnal, Vol. 33, No. 7,
2007, pp. 492-512.}

\title{\bf High Mass X-ray Binaries and Recent Star Formation History of the Small Magellanic Cloud}

\author{P. E. Shtykovskiy$^{1,2 \,*}$, M. R. Gilfanov$^{2,1}$}

\affil{
$^{1}$Space Research Institute, Profsoyuznaya str. 84/32, Moscow 117997, Russia\\
$^{2}$MPI for Astrophysik, Karl-Schwarzschild str. 1, Garching, 85741, Germany}

\sloppypar
\vspace{2mm}
\noindent

We study the relation between high-mass X-ray binary (HMXB) population and recent
star formation history (SFH) for the Small Magellanic Cloud (SMC). Using archival optical SMC
observations, we have approximated the color-magnitude diagrams of the stellar population by model
stellar populations and, in this way, reconstructed the spatially resolved SFH of the galaxy over the past
100 Myr.We analyze the errors and stability of this method for determining the recent SFH and show that
uncertainties in the models of massive stars at late evolutionary stages are the main factor that limits its
accuracy. By combining the SFH with the spatial distribution of HMXBs obtained from XMM-Newton
observations, we have derived the dependence of the HMXB number on the time elapsed since the star
formation event. The number of young systems with ages  10 Myr is shown to be smaller than the
prediction based on the type-II supernova rate. The HMXB number reaches its maximum $\sim$20--50~Myr
after the star formation event. This may be attributable, at least partly, to a low luminosity threshold in the
population of X-ray sources studied, Lmin$\sim10^{34}$~erg/s. Be/X systems make a dominant contribution to
this population, while the contribution from HMXBs with black holes is relatively small.


\vfill

{$^{*}$ E-mail: pavel@hea.iki.rssi.ru}
\newpage
\thispagestyle{empty}
\setcounter{page}{1}

\section*{INTRODUCTION}

High-mass X-ray binaries (HMXBs) are close
binary systems in which the compact object (a black
hole or a neutron star) accretes matter from an early-type
massive star. Because of the short lifetime of
the donor star, they are closely related to recent star
formation and, in the simplest picture, their number
should be roughly proportional to the star formation
rate of the host galaxy. Indeed, Chandra observations
of nearby galaxies suggest that, to the first approximation,
the HMXB luminosity function follows a universal
power law whose normalization is proportional
to the star formation rate (SFR) of the host galaxy
(Grimm et al. 2003).

On the other hand, obvious considerations based
on the present view of the evolution of binary systems
suggest that the relation between HMXB population
and star formation should be more complex
than a linear one. There is also experimental evidence
for this. For example, previously (Shtykovskiy and
Gilfanov 2005a), we showed that the linear relation
between the number of HMXBs and the SFR cannot
explain their spatial distribution over the Large
Magellanic Cloud (LMC), because their number does
not correlate with the H$_{\alpha}$ line intensity, a well-known
SFR indicator. The largest number of HMXBs is
observed in the region of moderate star formation
LMC 4, while they are virtually absent in the most
active star-forming region in the LMC, 30 Dor. Previously
(Shtykovskiy and Gilfanov 2005a), we suggested
that this discrepancy could arise from the dependence
of the HMXB number on the time elapsed
since the star formation event. Indeed, the age of the
stellar population in 30 Dor is $\approx1-2$~Myr, which is
not enough for the formation of compact objects even
from the most massive stars and, accordingly, for the
appearance of accreting X-ray sources. At the same
time, the characteristic age of the stellar population
in LMC 4, $\approx10-30$~Myr, is favorable for the formation
of an abundant HMXB population. Thus, on the spatial
scales corresponding to individual star clusters,
the linear relation between the HMXB number and
the instantaneous SFR does not hold and the recent
star formation history (SFH) on time scales of the
order of the lifetime of the HMXB population, i.e.,
$\sim2-100$~Myr, should be taken into account. Obviously,
the number of active HMXBs at a certain time
is determined by the total contribution from systems
of different ages according to the dependences of the
star formation history SFR(t) and a certain function
$\eta_{HMXB}(t)$ describing the dependence of the HMXB
number on the time elapsed since the star formation
event. The universal relation N$_{HMXB}=A\times$SFR on
the scales of galaxies results from the spatial averaging
of $\eta_{HMXB}(t)$ over star-forming regions of different
ages.

The Small Magellanic Cloud (SMC) is an ideal
laboratory that allows these and other aspects of
HMXB formation and evolution to be studied. Indeed,
owing to its appreciable SFR and small distance
(60 kpc), there are dozens of known HMXBs in
it. On the other hand, the SMC proximity makes
it possible to study in detail its stellar population
and, in particular, to reconstruct its SFH. Another
peculiarity of the SMC, namely, its low metallicity,
makes it potentially possible to study the effect of
the heavy-element abundance on the properties of
the HMXB population. In this paper, we use XMM-Newton
observations of the SMC (Shtykovskiy and
Gilfanov 2005b) and archival optical observations
(Zaritsky et al. 2002) to analyze the relation between
the number of HMXBs  and the recent SFH of the galaxy.
Our goal is to derive the dependence of the HMXB
number on the time elapsed since the star formation
event.

\section{EVOLUTION OF THE HMXB POPULATION
AFTER THE STAR FORMATION EVENT} 
\label{sec:hmxbevol}

To describe the evolution of the HMXB population,
let us introduce a function $\eta_{HMXB}(t)$ that describes
the dependence of the number of observed
HMXBs with luminosities above a given value on the
time t elapsed since the star formation event normalized
to the mass of the formed massive stars:
\begin{eqnarray}
\eta_{HMXB}(t)=\frac{N_{HMXB}(t)}{M(>8M_{\odot})}
\label{eq:etahmxbteor1}
\end{eqnarray}
where M($>$8~M$_{\odot}$)  is the mass of the stars more massive
than 8~M$_{\odot}$ formed in the star formation event
and N$_{HMXB}(t)$ is the number of HMXBs with luminosities
exceeding a certain threshold. The luminosity
of $10^{34}$~erg/s that corresponds to the sensitivity
achieved by XMM-Newton in the SMC observations
is taken as the latter.

Obviously, the function $\eta_{HMXB}(t)$ is non-zero only
in a limited time interval. Indeed, the first X-ray binaries
appear only after the formation of the first black
holes and/or neutron stars. The lifetimes of the stars
that explode as type II supernovae (SNe II) to produce
a compact object lie in the interval from $\approx2-3$~Myr
for the most massive stars, $\approx100$~M$_{\odot}$, to 
$\approx40$~Myr for
stars with a mass of $\approx$8~M$_{\odot}$, the least massive stars
capable of producing a compact object. In this picture,
it would be natural to expect the X-ray binaries
in which the compact object is a black hole to appear
first and the (probably more abundant) population of
accreting neutrons stars to appear next.

On the other hand, the HMXB lifetime is limited
by the lifetime of the companion star. Since the least
massive companion stars observed during an active
X-ray phase have a mass of $\approx6 M_{\odot}$, this lifetime is
$\sim60$~Myr for a single star when the peculiarities of the
stellar evolution in binary systems are disregarded.
Given the mass transfer from the more massive star
to the future donor star, this lifetime can be slightly
modified. This also includes the X-ray source stage
proper with characteristic time scales much shorter
than those considered above,  $\sim10^3-10^6$~yr, depending
on the type of the companion star and the binary
parameters.

Obviously, the function  $\eta_{HMXB}(t)$ must be closely
related to the rate of SNe II  $\eta_{SNII}(t)$ producing a
compact object. To the first approximation, the relation
may be assumed to be linear:
\begin{eqnarray}
\eta_{HMXB}(t)= A\cdot\eta_{SNII}(t)
\label{eq:etahmxbteor0}
\end{eqnarray}
The supernova rate can be easily determined from
the stellar mass--lifetime relation (Schaller et al.
1992) and the initial mass function (IMF), which
below is assumed to be a Salpeter one in the range
0.1--100M$_{\odot}$. Note that the IMF shape in the range
of low masses is unimportant for us, since all of the
relations are eventually normalized to the mass of
massive stars with M$>$8~M$_{\odot}$. The normalization in
Eq. (2) can be calculated using the N$_{HMXB}$--SFR
calibration from Grimm et al. (2003). This relation
was derived from Chandra observations of nearby
galaxies and corresponds to the time integral of the
function  $\eta(t)$):
\begin{equation}
\int \eta_{HMXB}(t) dt=\frac{N_{HMXB}(L_X>L_{X,min})}{SFR}
\label{eq:etahmxbteornorm}
\end{equation}
As the limits of integration in Eq. (3), we choose 2 and
40 Myr in accordance with the above reasoning.

In what follows, we will compare the experimental
dependence $\eta_{HMXB}(t)$ obtained from X-ray and optical
SMC observations with predictions of the simple
model specified by Eqs. (2) and (3). Clearly, Eq. (2)
is based on the assumption that the X-ray phase
comes immediately after the formation of a compact
object, i.e., it disregards the evolution of the companion
star in the binary system. A more rigorous
description of the HMXB evolution requires resorting
to population synthesis models (see, e.g., Popov and
Prokhorov 2004; Belczynski et al. 2005), which is 
outside the scope of this paper. On the other hand, the
experimental dependence $\eta_{HMXB}(t)$, whose derivation
is the goal of this paper, can be used by the creators
and users of population synthesis models to test and
calibrate these models.

\subsection{Experimental Determination of the Function $\eta_{HMXB}(t)$}

The number of HMXBs observed in a spatial region
X at time t is a convolution of the function
$\eta_{HMXB}(t)$ with the star formation history SFR(t,X)
in this region:
\begin{equation}
N_{HMXB}(t,X)=\int SFR(t-\tau,X)\eta_{HMXB}(\tau) d\tau.
\label{eq:etahmxbobs}
\end{equation}
Solving the inverse problem formulated by this equation,
we can impose constraints on the dependence
 $\eta_{HMXB}(t)$ from observations. This requires the following:

\begin{enumerate}
\item Identifying the HMXB population in the
galaxy.
\item Reconstructing the spatially-resolved star formation
history SFR(t,X). Obviously, we need only
the recent SFH from the current time to the time
in the past corresponding to the maximum HMXB
lifetime (i.e., $\sim 50-100$~Myr).
\item Solving the inverse problem formulated by
Eq. (4) given $N_{HMXB}(X)$ and SFR(t,X) for a large
set of regions X.
Obviously, a galaxy with a rich HMXB population
and a SFH that changes significantly from place to
place is required to perform this procedure. Because
of its proximity and appreciable SFR, the SMC is one
of the most natural candidates for such a galaxy. This
paper is structured as follows. We describe the SFH
reconstruction technique and apply it to the SMC,
solve the inverse problem given by Eq. (4), and find
the function $\eta_{HMXB}(t)$ for HMXBs in the SMC. Next,
we discuss the results obtained and summarize our
conclusions.
\end{enumerate}

\section{THE STAR FORMATION HISTORY IN THE
SMALL MAGELLANIC CLOUD}
\label{sec:sfh}

To reconstruct the SFH, we will use a method
based on the analysis of color-magnitude diagrams
(see, e.g., Gallart et al. 2005). This method uses the
fact that stars of different ages (and metallicities)
occupy different positions in the color-magnitude
diagram. The SFH can be determined by comparing
the distributions of stars in it with predictions
of stellar evolution models. Applying this method
requires optical photometry at least in two bands.
There are several realizations of this method; one of
the most commonly used realizations was described
by Dolphin (1997), Aparicio et al. (1997), and Dolphin
(2002) and consists of the following steps:
\begin{enumerate}
\item Generating synthetic color-magnitude diagrams
in the required ranges of metallicities and ages
on the basis of stellar evolution models. Each diagram
is the probability distribution in color-magnitude
space for a coeval model stellar population.
\item Correcting the synthetic diagrams for the incompleteness
and photometric errors. Allowance for
the interstellar extinction and the distance to the
galaxy.
\item Approximating the observed color-magnitude
diagrams by a linear combination of the derived synthetic
models. Estimating the uncertainties of the
solution.
\end{enumerate}

Because of their proximity, the Magellanic Clouds
are attractive objects for star formation studies. It is
not surprising that a number of papers are devoted
to the SFH in them (see, e.g., Holtzman et al. 1999;
Dolphin 2000). In particular, note the paper by Harris
and Zaritsky (2004), who reconstructed the spatially
resolved SFH of the SMC. However, in all of these
studies, the star formation was considered in a wide
range of ages, with the emphasis being inevitably
on time scales of $\sim$~Gyr. In contrast, we are interested
in the SFH for the youngest stellar population.
As will be shown below, its reconstruction has several
peculiarities that have escaped attention previously.
Therefore, we adapted the SFH reconstruction
method to meet the requirements of our problem by
concentrating on the time interval 0--100~Myr.

\subsection{Synthetic Color-Magnitude Diagrams}
\label{sec:synthcmd}

The first step in generating synthetic color-magnitude
diagrams is to choose the model isochrones
that define the region occupied by a coeval stellar
population. In what follows, we use the isochrones
from Girardi et al. (2002) (the so-called ``Padova
isochrones'') covering wide ranges of ages (log t =
6.60--10.25), metallicities (Z = 0.0001--0.03), and
masses (0.15--70~M$_{\odot}$). All model calculations are performed
for the color-magnitude diagrams in (U--B, B) and (B--V , V) spaces.

The theoretical isochrones relate the mass of a
star of a certain age to its position in the diagram.
Therefore, the probability of filling some region in
it can be easily determined from the corresponding
mass interval M$_i$--M$_{i+1}$ and the IMF, which below is
assumed to be a Salpeter one:
\begin{equation}
p(M_{i},M_{i+1})=\frac{M_{i+1}^{-\Gamma}-M_{i}^{-\Gamma}}{M^{-\Gamma}_{max}-M^{-\Gamma}_{min}},
\label{eq:occprob}
\end{equation}
where $\Gamma=1.35$, $M$ is the initial mass of the star,
$M_{min}$=0.1~$M_{\odot}$, and M$_{max}$=100~$M_{\odot}$. Note that the
IMF deviations from the Salpeter one in the range
of low masses affect only the normalization of the
derived SFH rather than its shape. This is because
we analyze the color-magnitude diagrams only for a
relatively massive stellar population. The SFH sensitivity
to the IMF deviations from the Salpeter one
in the range of high masses is discussed in the Section
``Checking the SFH Reconstruction Procedure.''
Equation (5) allows us to calculate the probabilities
of filling various regions in the color-magnitude diagram
that are needed to fit the observations by a
model. This is convenient to do using model photometry
generated by the Monte Carlo method. The
total number of model stars must be large enough
to minimize the contribution from Poisson noise. In
our case, the number of stars is $>10^5$ per isochrone
(which corresponds to 10$^8$ stars in the mass range
0.1--100~M$_{\odot}$).

However, before generating model photometry, we
must make several more steps, including the choice of
an age range, an age step, ametallicity range, a binary
fraction and isochrone interpolation. These steps are
considered below.

First, we found that the isochrones need to be
interpolated. Indeed, the magnitude difference at adjacent
points can be 0.5. Therefore, we perform a
linear interpolation of the magnitudes in such a way
that the magnitude step does not exceed 0.01.

In choosing an age interval and its binning, we will
keep in mind that we are interested only in the recent
star formation. This allows us to exclude the old population
from our analysis and, thus, to avoid problems
related to the incompleteness of the optical catalog
at faint magnitudes (see the Subsection ``Binning
of Color-Magnitude Diagrams''). Below, we reconstruct
the SFH in the time interval log t = 6.6--8.0.
We also include the isochrones in the time interval
log t = 8.0--8.6 in our model to avoid the distortion of
the solution at log t$\leq$8.0 due to the older population.
Initially, the time step in the isochrones is $\Delta\log(t)=0.05$. 
The simple tests show that this resolution
is excessive in terms of the photometry used (see
the Subsection ``Optical Photometry''). Therefore, we
combine the isochrones into groups, each with 3--4
isochrones, thereby obtaining the time step $\Delta\log(t)=0.2$.

The binary fraction is also important in generating
model stellar populations, since the binary stars in the
color-magnitude diagram will appear as single stars
with distorted photometry. As the binary fraction,
we use the standard value of $f_{binary}=0.5$. Following
Harris and Zaritsky (2004), we will assume that the
mass of the companion star is taken from an independent
Salpeter IMF. The influence of these assumptions
on the derived SFH is discussed in the Section
``Checking the SFH Reconstruction Procedure.''

\subsubsection{Metallicity.}
\label{sec:metallicity} 

The heavy-element abundance is an
important parameter in the evolution of a star. The
positions of stars with different metallicities in the
color-magnitude diagram will differ almost at all evolutionary
stages. For example, since an increase in
metallicity is accompanied by an increase in opacity,
it causes the main sequence to be displaced toward
the less bright and cooler stars. However, metallicity
plays the most important role at the final stages of
stellar evolution. For instance, the position of a star
in the color-magnitude diagram for (super)giants
depends critically on the heavy-element abundance.
This can give rise, for example, to partial degeneracy
between age and metallicity for red supergiants.
Therefore, choosing the isochrones with the proper
metallicity (or metallicity range) is very important for
reconstructing the SFH.

The metal abundance in the Magellanic Clouds
is known to be low. For example, the metallicity of
the interstellar medium in the SMC is 0.6 dex lower
than that of the local medium in our Galaxy (Russell
and Dopita 1992). Note also that the SMC metallicity
has gradually increased with time due to continuous
star formation. Therefore, a self-consistent
description of the SFH should take into account
the spread in metallicity. Several attempts have been
made to describe quantitatively the heavy-element
enrichment history of the SMC. The typical metallicities
lie in the range from
$[Fe/H]\approx-1.25$ (Z$\approx0.001$) for the old population to 
$[Fe/H]\approx-0.5$ (Z$\approx0.006$) for the young population (see, e.g., Pagel and
Tautvaisiene 1998). Since we are interested in the
recent star formation in the SMC, a component relatively
rich in heavy elements is expected to dominate
among the stellar population used in the calculations.
However, a spread in metallicity exists even for
the young population (see, e.g., Harris and Zaritsky
2004; Maeder et al. 1999; and references therein).
Therefore, to choose the metallicity suitable for the
spatial regions used to reconstruct the SFH, we visually
compare the observed color-magnitude diagrams
with the model isochrones. The effect of the
heavy-element abundance is most pronounced for
the red supergiant branch. Note that the isochrones
in the region of red supergiants in the range  $Z\sim 0.004-0.008$ 
under consideration do not intersect,
i.e., there is no degeneracy between metallicity and
age. We found that the locations of the red supergiant
branches in most regions are satisfactorily described
by the Z = 0.004 isochrones. However, in one region,
the (B--V, V) diagram is described better by Z =
0.008, while Z = 0.004 is more suitable for the (U--B, B)
 diagram. Below, we use Z = 0.004 everywhere,
except for this region where Z = 0.008 is used. We
also analyze the dependence of our results on the
chosen metallicity (see below).

\subsubsection{Interstellar extinction and distance.}
\label{sec:synthcmd1}

 The derived
synthetic diagrams should also be corrected for
the interstellar extinction and the SMC distance. As
the distance modulus for the SMC, we take m--M =
18.9 (Westerlund 1997), corresponding to a distance
of D$\approx$60~kpc.

Zaritsky et al. (2002) showed that the extinction
for the stellar population in the SMC changes from
region to region and differs for hot and cool stars
and obtained the distributions of extinction for different
regions (http://ngala.as.arizona.edu/dennis/smcext.html). 
We correct the synthetic photometry
using these distributions just as was done by
Harris and Zaritsky (2004). For the young stars
($logt < 7.0$), we take the distribution of extinction
corresponding to hot stars. For the older population,
the distribution of extinction for hot stars is used
only for the population fraction $f=1-0.5\cdot(\log(t)-7)$,
while the fraction 1--f of stars exhibit extinction
corresponding to cool stars. Finally, all of the stars
older than 1 Gyr have the distribution of extinction
corresponding to cool stars.

\subsubsection{Photometric errors and completeness.}
\label{sec:photoerror}

 The synthetic color-magnitude diagrams should take into
account the photometric errors and the incompleteness
of the optical catalog at low fluxes. The most
important source of errors is the telescope's limited
resolution. Clearly, the resolution-related photometry
distortions depend on the spatial density of stars and
will be at a maximum where this density is high.
Artificial star tests -- reconstructing the photometry
of model stars placed on real images through
the standard procedures used in compiling a real
catalog -- are a standard method of solving this problem.
The subsequent comparison of the reconstructed
photometry with the model one allows the distortions
produced by this factor to be estimated as a function
of the spatial density of stars. Obviously, this requires
input optical data. In addition, there are factors whose
contribution is much more difficult to estimate quantitatively
(e.g., the systematic uncertainties in the
calibration). Analysis of the star catalog used shows
that these are actually present (see the Subsection
``Optical Photometry'').
If the photometric errors are moderately large, then
the problem of photometric errors can be solved by
choosing a special binning of the color-magnitude
diagram, more specifically, using a grid with wider
color and magnitude intervals than the characteristic
photometry distortions. Since we are interested only
in the recent SFH, we can also exclude faint stars
for which the problem of photometric errors is more
serious. Excluding faint stars also solves the problem
with the incompleteness of the catalog. On the other
hand, it is clear that we cannot make the cells in
the color-magnitude diagram too large, because this
can give rise to additional degeneracy in the solution.
Since there were no input optical data for the
SMC at our disposal, we chose the second path --
optimizing the binning of the color-magnitude diagram
(for more detail, see the Subsection ``Binning of
the Color-Magnitude Diagram'').

\subsection{SFH reconstruction.}
\label{sec:sfhfit}

 Using the synthetic photometry
obtained, we can approximate the observed
distribution of stars in the color-magnitude diagram
($n_i$) by linear combinations of model stellar populations
($A_{i,j}$ ):
\begin{equation}
n_i=\sum_j A_{i,j}\times x_j,
\label{eq:cmdapprox}
\end{equation}
where i is the cell number in the diagram and j is the
time interval number. The amplitudes $x_j$ minimizing
the discrepancy  $\Vert Ax-n\Vert$ are the sought-for SFH.

Since the problem in question is ill-conditioned,
we use an iterative Lucy-Richardson method (Lucy 1974) 
for its solution. Using the initial approximation
to the solution, this procedure calculates
a vector that approaches the maximum likelihood
solution with increasing number of iterations. The
solution after iteration i is regularized in the sense
that the method retains non-negativity of the initial solution
and that it is smoother than themaximum likelihood
solution. An important feature of the method
is the choice of a stopping criterion (Lucy 1994) --
the number of iterations giving an optimum solution.
Obviously, the stopping criterion is determined by the
character of the problem. For example, at low noise
in the input data, the maximum likelihood solution is
close to the true one, while in the reverse situation
with large errors, fitting the data with a high accuracy
is equivalent to attempting to describe the noise.
Below, we define a stopping criterion suitable for our
problem by reconstructing the SFH of a model stellar
population and studying the behavior of the likelihood
function L depending on the number of iterations (see
below):
\begin{equation}
L=\sum_i (\mu-N_i\cdot \ln\mu).
\label{eq:statistics}
\end{equation}

\begin{figure*}
\includegraphics[width=1.0\textwidth,clip=true]{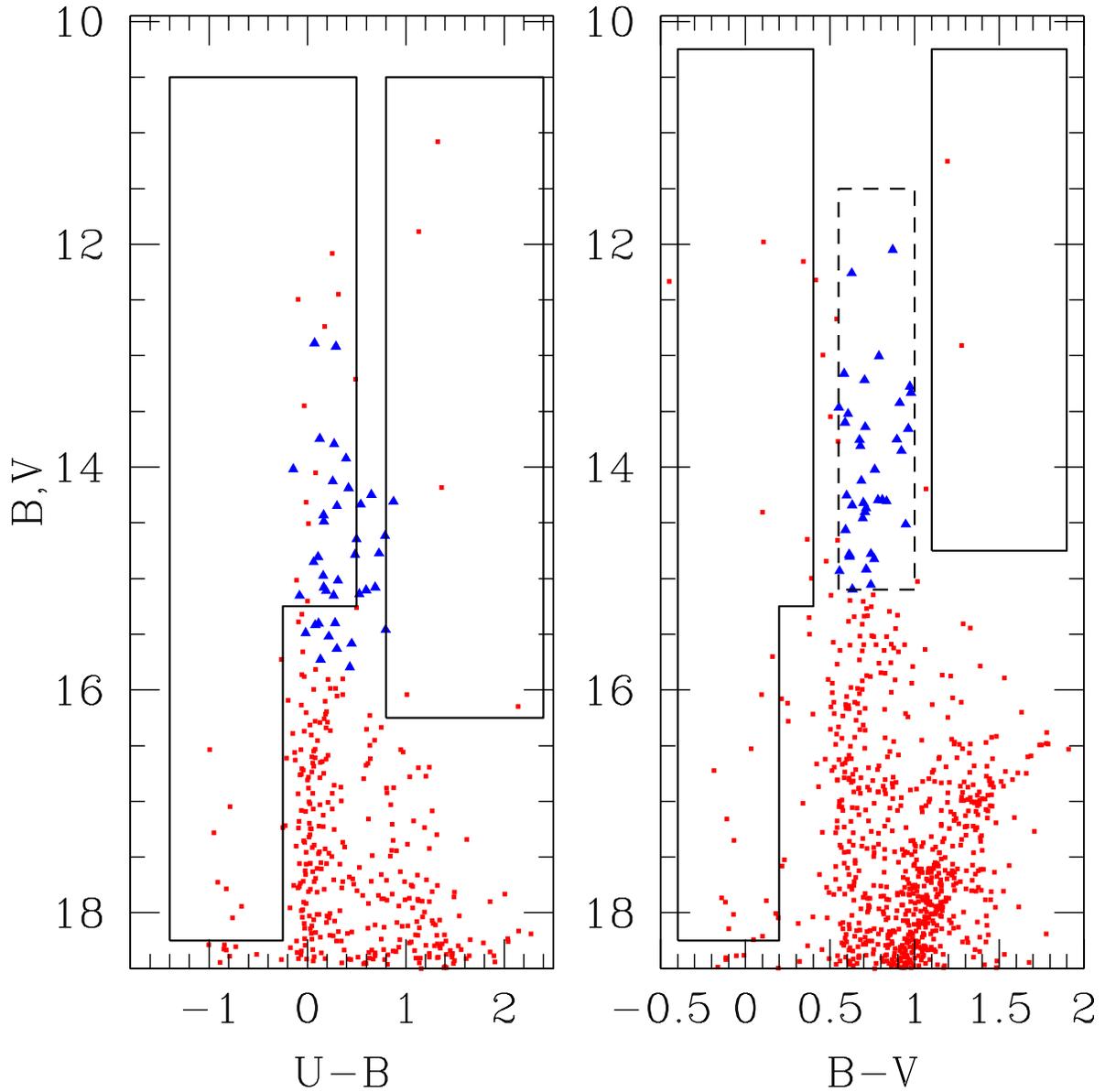}
\caption{Color--magnitude diagram for the extreme MCPS fields 
illustrating the smallness of the contribution from Galactic
foreground stars and the efficiency of the method of their rejection. 
The rectangles (solid line) indicate the regions used to
reconstruct the SFH. The dashed line in the (B--V, V) diagram denotes 
the region that was used to identify foreground stars.
The stars that fell into this region are marked by triangles in both 
diagrams. They were assumed to belong to the Galaxy and
were excluded from the analysis in the (U--B, B) diagram.
}
\label{fig:cmdgridfg}
\end{figure*}

\subsubsection{Optical photometry.}
\label{sec:optcat} 

As the stellar population
photometry necessary to reconstruct the SFH,
we used the Magellanic Clouds Photometric Survey
(MCPS) catalog for the SMC (Zaritsky et al.
2002). To reconstruct the SFH, we use the (U--B, B)
and (B--V , V) diagrams. The catalog also presents
I-band photometry, i.e., the additional (V--I, I)
diagram could be used. However, we found that the
I magnitude is often absent (I = 0) for bright stars,
with the most significant loss of photometry being
observed among the red supergiants. Since the latter
play an important role in reconstructing the recent
SFH, we decided to exclude the (V--I, I) diagram
from our analysis.

Note also the problem with the U photometry of
the catalog. As described in Zaritsky et al. (2004),
Zaritsky et al. (2002) corrected the U--B color using
the photometry from Massey (2002) calibrated (in the
initial version) from faint dwarfs. In addition, Zaritsky
et al. (2002) replaced part of the photometry for bright
stars with the photometry from Massey (2002). As
a result, the U--B color for blue supergiants may
be unreliable. This is clearly seen in the (U--B, B)
diagram as the displacement of the blue supergiant
sequence by $\sim0^m.3$ relative to the model (see, however,
the ``Section Checking the SFH Reconstruction
Procedure'' for a discussion of the reliability of stellar
evolution models for supergiants). Our binning of the
color-magnitude diagram into large cells allows the
effect of this kind of uncertainties to be minimized.
Therefore, we expect this problem to be not critical in
our procedure. Another problem described by Harris
and Zaritsky (2004), more specifically, the need for
displacing the B-V color by $0^m.1-0^m.2$ in some regions,
will not affect strongly our results for the same
reason.
For test purposes, we also used the OGLE catalog
(Udalski et al. 1998) containing B, V, I photometry,
but covering only part of the SMC.

\begin{figure*}
\includegraphics[width=1.0\textwidth,clip=true]{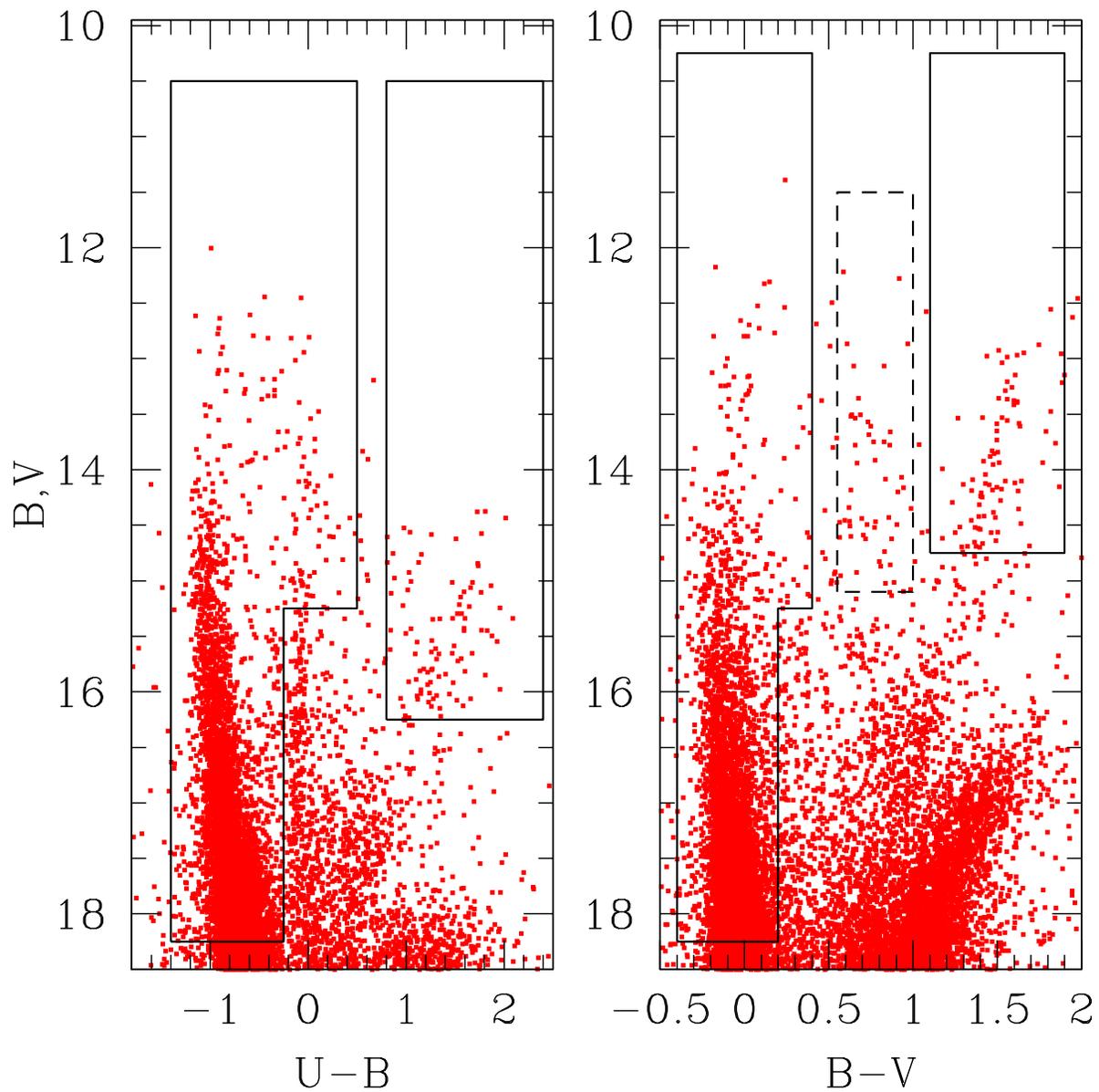}
\caption{Color-magnitude diagram for one of the SMC fields used to reconstruct 
the SFH. The notation is the same as that in Fig. 1.}
\label{fig:cmd}
\end{figure*}

\subsubsection{Contribution from foreground stars.}
\label{sec:foreground}

 Obviously, the catalog of stars toward the SMC also
contains Galactic stars that can introduce distortions
into the color-magnitude diagrams. To estimate their
contribution, we constructed the color-magnitude
diagrams for 10 outermost MCPS fields (each with an
area of 12$\arcmin\times$12$\arcmin$) where the contribution from SMC
stars is at a minimum (Fig. 1). When comparing the
densities of stars in Figs. 1 and 2, we should take
into account the fact that the total area of the fields
used to construct the diagram for Galactic foreground
stars shown in Fig. 1 is approximately twice the
area of the sky used to construct the diagram in
Fig. 2 (equal to the area of the XMM-Newton field
of view). Obviously, the contribution from Galactic
foreground stars is negligible in most of the color-magnitude diagram.
 The only region where Galactic
stars can introduce noticeable distortions is the blue
supergiant branch in the (U--B, B) diagram in the
range of colors near U--B = 0. However, as we see
from Fig. 1, they are easily identified in the (B--V, V) diagram, 
since they are separated from both blue
and red supergiants in it. This forms the basis for our
foreground star rejection algorithm. We determined
the region in the (B--V, V) diagram that, on the one
hand, includes most of the Galactic foreground stars
superimposed on the SMC blue supergiant branch in
the (U--B, B) diagram and, on the other hand, the
contribution from SMC stars to it is negligible. This
region is highlighted by the dashed line in Figs. 1
and 2. All of the stars lying in this region (they are
marked by triangles in both diagrams in Fig. 1) are
then excluded from the analysis in both (B--V, V)
and (U--B, B) diagrams.

\subsubsection{Binning of the color-magnitude diagrams.}
\label{sec:cmdgrid}

 To compare the model color-magnitude diagrams with
the observations, we must specify their binning. There
are two approaches to this problem -- uniform and
more complex grids. Whereas the former is more objective,
the latter makes it possible to avoid problems
related to the photometric errors and uncertainties
in the stellar evolution. In any case, the choice of
a grid must take into account the existence of extended
structures corresponding to long stages of
stellar evolution in the diagram. The main sequence
and blue and red supergiants are most important in
determining the recent SFH.

Since the core hydrogen burning is the longest
phase of stellar evolution, the number of stars on
the main sequence is at its maximum and the latter
plays a major role in determining the SFH. In principle,
the SFH can be reconstructed based only on
the main sequence, without invoking other stages of
stellar evolution (see, e.g., Dohm-Palmer et al. 1997).
However, this method places heavy demands on the
photometric accuracy, because the blue supergiants
are close to the upper part of the main sequence
(Figs. 2 and 3).

\begin{figure*}
\includegraphics[width=1.0\textwidth,clip=true]{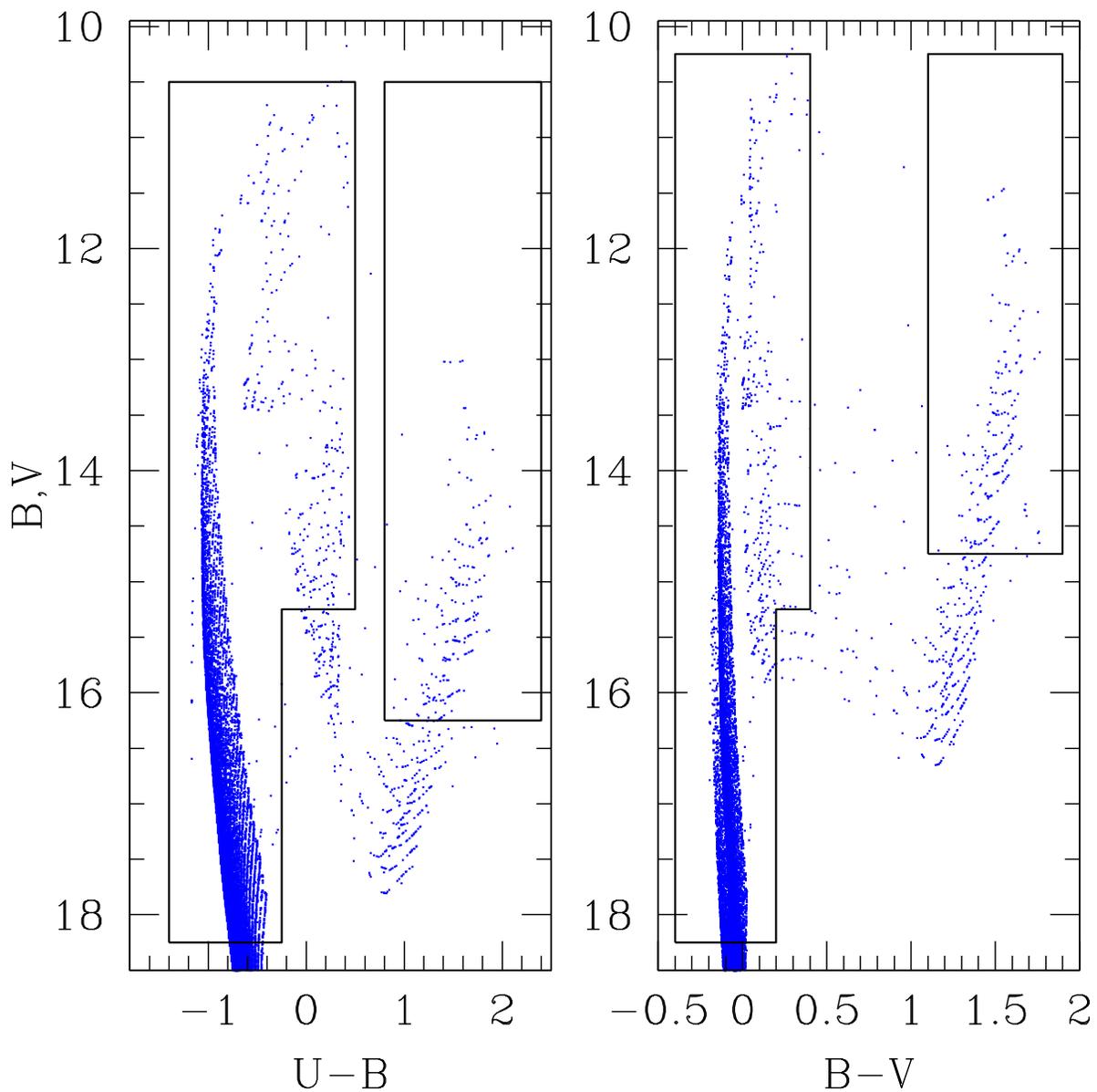}
\caption{Synthetic color-magnitude diagram. The rectangles denote the regions 
used to reconstruct the SFH. The SFH of a model population obeys the law 
dM/d log t = const in the time interval log t = 6.6--8.0 and is zero outside this interval.
}
\label{fig:cmdgrid}
\end{figure*}

The supergiant branches are important regions in
the color-magnitude diagram and complement the
main sequence when reconstructing the SFH. However,
whereas the evolution of a main-sequence star
has been studied well, the evolution of supergiants
is more uncertain. This is because the supergiants
are very sensitive to such aspects of the model as
mass loss, convection, etc. Uncertainties in the latter
can strongly affect the manifestation of supergiants
in the color-magnitude diagram. The best known
outstanding problem here is the blue-to-red supergiant
ratio (B/R). As was shown by Langer and
Maeder (1995), there are no stellar evolution models
that are capable of explaining self-consistently the
dependence of B/R on metallicity in a wide range of
the latter (see also Gallart et al. 2005).

The region that we use to compare the observations
with the model consists of two strips: one covers
the main sequence and the blue supergiant branch
and the other covers the region of red supergiants (see
Fig. 3). The width of each strip is taken to be much
larger than the scatter in photometry, which also allows
the effect of uncertainties in the stellar evolution
models to be reduced (for more detail, see the Section
``Checking the SFH Reconstruction Procedure''). At
the same time, it is small enough for the contribution
from Galactic foreground stars to be at a minimum
(Fig. 1). Since each strip has only one color interval,
using this grid is equivalent to simultaneously fitting
two luminosity functions. The scatter in magnitude
is less important than the scatter in color, because
all of the features in the color-magnitude diagram
are elongated along the magnitude axis. Its effect
is equivalent to convolving the SFH with the function
defined by the distribution of photometric errors.
We take dm = 0.25 as the width of the magnitude
interval.

The magnitude threshold for the main sequence
was chosen to be V$_{lim}$=18.25 and B$_{lim}$=18.25.
This allowed us to avoid problems with the incompleteness
of the catalog and with the photometry
distortion through the superposition of stars. Indeed,
for the MCPS catalog, the completeness is large for
$\sim20^m$ stars and the magnitude errors (including the
star superposition effect) for bright stars are smaller
than the chosen width of the color and magnitude
intervals (Zaritsky et al. 2002). We used a higher
threshold for red supergiants to avoid the contribution
from the old low-metallicity stellar population.

\subsubsection{Uncertainty of the solution.}
\label{sec:sfhfit3}

To estimate the
statistical uncertainty in the reconstructed SFH, we
analyzed the stability of our solution to Poisson noise
in the number of stars by the bootstrap method. We
calculated the expected number of stars in each cell
in the color-magnitude diagram from our solution.
Next, we drew their realization by assuming a Poisson
distribution for the number of stars, which was
then used as input data in the SFH reconstruction
code. This procedure was repeated many times and
the rms scatter of the solutions obtained was taken as
the error.

\subsection{Checking the SFH Reconstruction Procedure.}
\label{sec:sfhval}

To check the SFH reconstruction procedure, we
performed a number of tests. First of all, to check
the general functioning of the algorithm and its implementation,
we reconstructed the SFHs for various
model stellar populations. Subsequently, we investigated
the adequacy of the stellar evolution models
and the accuracy with which the observed color-magnitude diagrams 
are approximated. Finally, we
analyzed the stability of the solution to photometric
errors and its sensitivity to various model parameters,
such as the metallicity, the binary fraction, and the
IMF slope.

\bigskip
{\it Reconstructing the SFH for a model stellar
population.} 

For the first test, we chose a model stellar
population whose SFH consists of several bursts
alternating with periods of quiescence. The number
of stars in the model population was close to
that observed in the SMC within the XMM-Newton
field of view. This test allows us to check the SFH
reconstruction procedure, the degeneracy between
adjacent time intervals, and to analyze the dependence
of the solution on the stopping criterion in the
Lucy-Richardson method. The results are presented
in Fig. 4, which shows the behavior of the likelihood
function depending on the number of iterations and
the model and reconstructed SFHs. We compare two
solutions -- one long before the saturation of the likelihood
function (200 iterations) and the other close to
its saturation (1000 iterations). Obviously, the latter
corresponds much better to the model.

The model SFH used in the first test is implausibly
complicated. As a more realistic example, we chose
the actual SFH obtained for one of the SMC fields
and used the model stellar population corresponding
to it as input data in our code. As we see from Fig. 4,
the model SFH is smoother in this case. As in the
previous case, the best solution is achieved close to
the saturation of the likelihood function.

Based on the results of these and other tests, we
concluded that the best solution is achieved near the
saturation of the likelihood function. In other words,
the problem has such a character (the number of stars
etc.) that the solution obtained requires no (or almost
no) significant regularization.

\begin{figure*}
\centerline{
\vbox{
\hbox{
\resizebox{0.40\hsize}{!}{\includegraphics{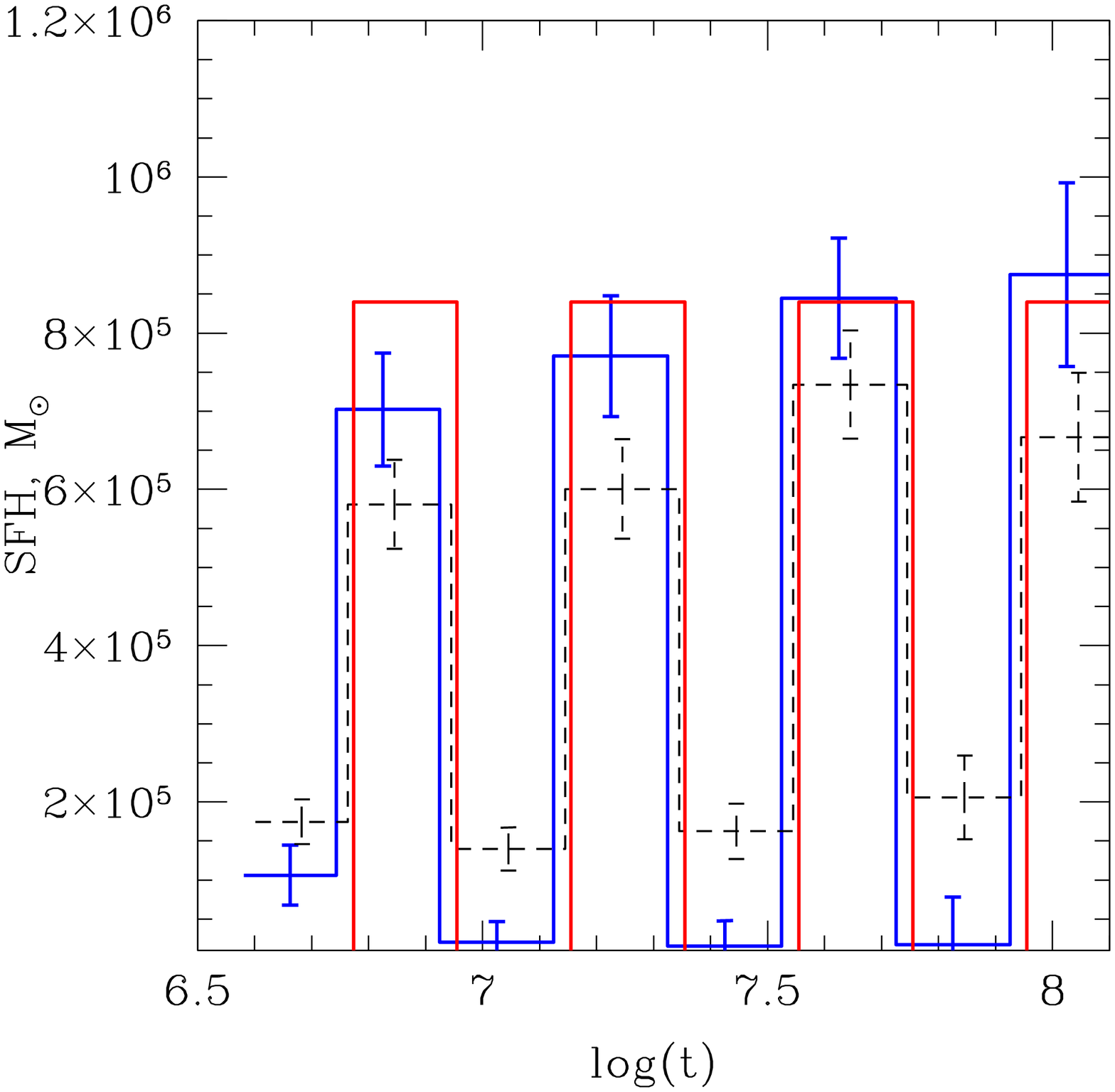}}
\resizebox{0.40\hsize}{!}{\includegraphics{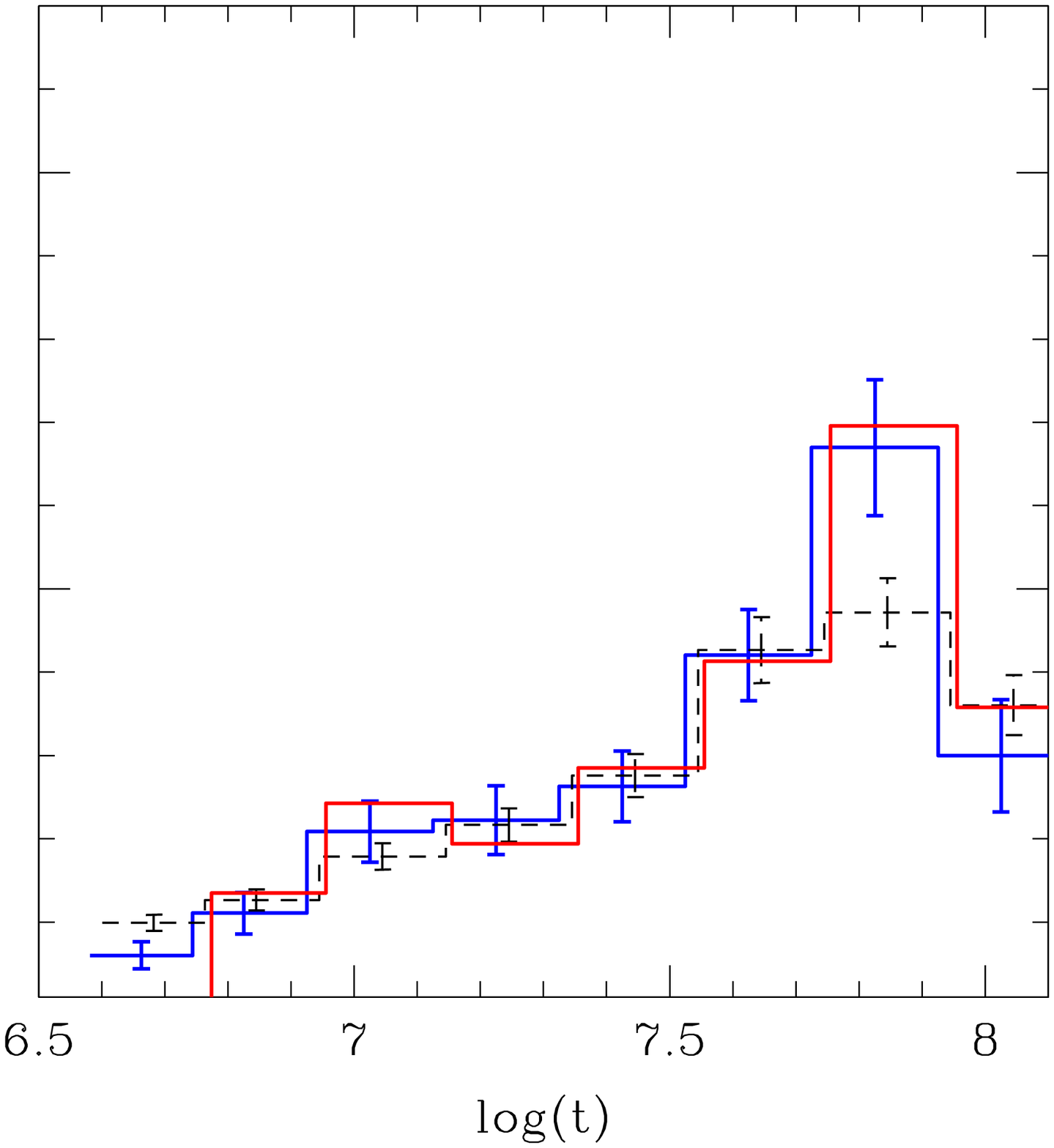}}
}
\hbox{
\resizebox{0.40\hsize}{!}{\includegraphics{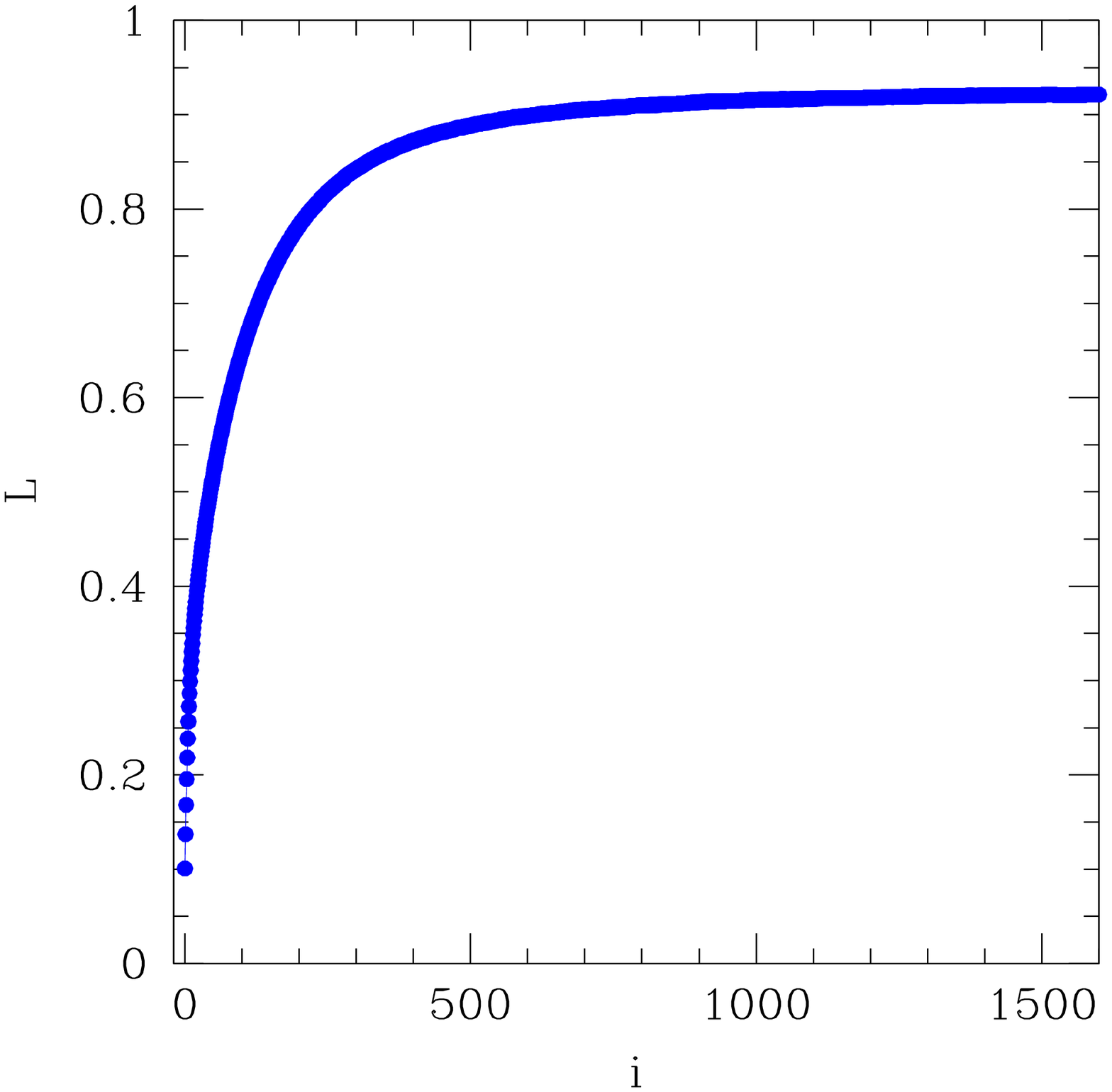}}
\resizebox{0.40\hsize}{!}{\includegraphics{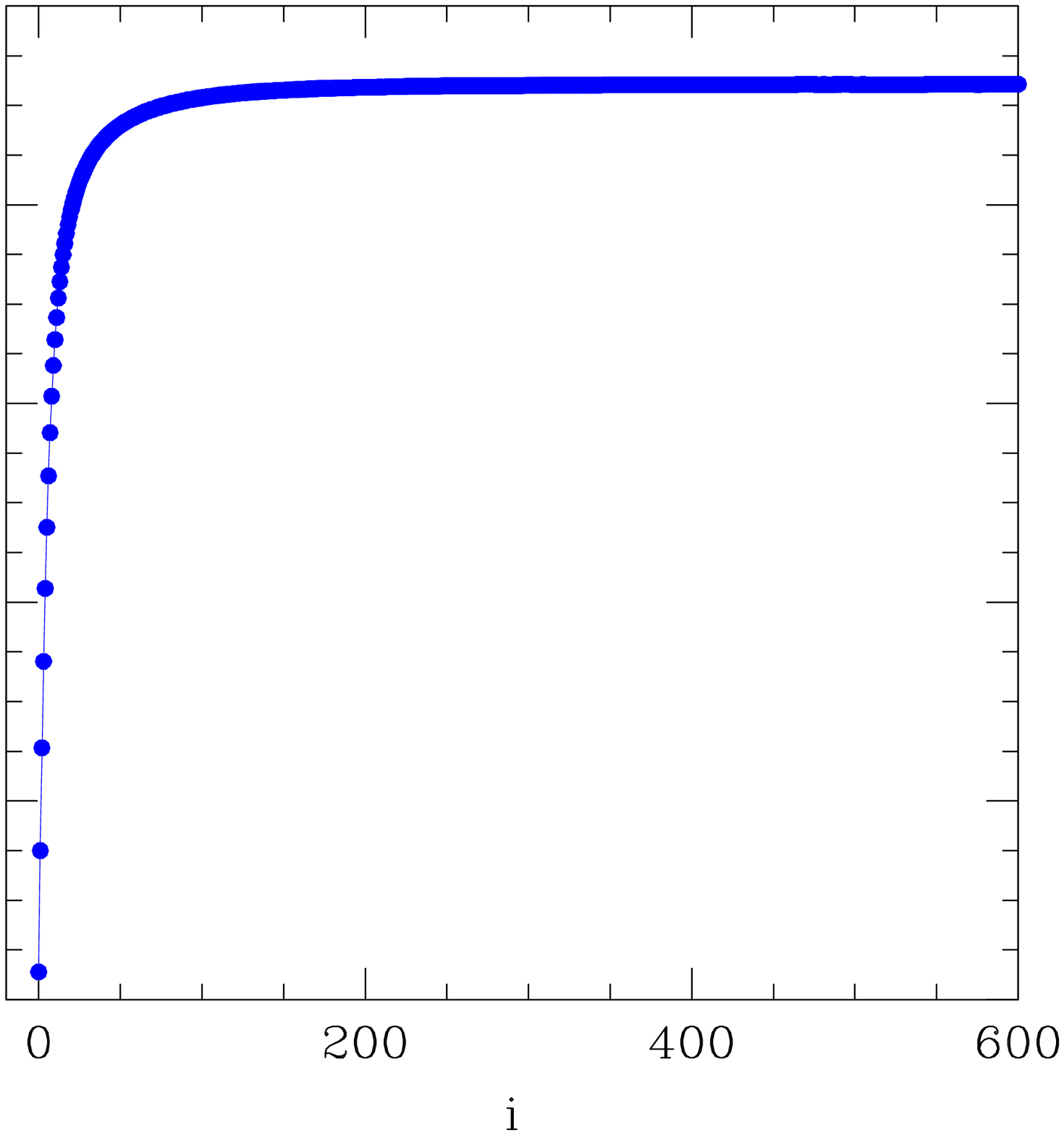}}
}
}
}
\caption{(a), (b) Reconstructed SFHs for model stellar populations. 
The mass formed in different time intervals is along the
vertical axis. The model SFH is indicated by the thick line without error bars.
 For clarity, the histograms were displaced
along the time axis. (c), (d) Behavior of the likelihood function for the solution 
depending on the number of iterations in the
Lucy-Richardson method. Model is a series of star formation events 
alternating with periods of quiescence (a) 
and the SFH in one of the SMC fields (b). The solid and
dashed histograms indicate the solutions obtained by the Lucy-Richardson method 
after 1000, 200 (a) and 250, 40 (b)
iterations, respectively. 
}
\label{fig:sfhsim}
\end{figure*}

\bigskip
{\it The adequacy of stellar evolution models.}

To verify the adequacy of the model isochrones, we
analyzed how well the model describes the color-magnitude diagram 
for the actual stellar population.
Figure 5 presents the model and observed luminosity
functions for themain sequence with blue supergiants
and for red supergiants summed over all of the SMC
fields used in this paper. These luminosity functions
correspond to the two regions shown in Fig. 2. We
see that the model agrees well with the observations
for faint magnitudes, but in the region of bright stars
 ($B,V\la 13.5$) the model prediction for the main
sequence and blue supergiants exceeds appreciably
the observations. As our tests showed, this excess
is related to blue supergiants -- the two clearly seen
features in the model luminosity function at V$\approx$13
and $\approx11.5$, which are much less pronounced in the
data, are unequivocally identified with them. This is
clear from an examination of the color-magnitude
diagram for the model population in the (U--B, B)
diagram in Fig. 3. The problem with the excess for the
brightest stars can be partly removed if the metallicity
is assumed to be Z = 0.008 for all fields. In this case,
however, the locations of the supergiant branches
in the color-magnitude diagram will be in poorer
agreement with the data. On the other hand, the luminosity
function for red supergiants is described well
by the model. Obviously, the discrepancy between
the data and the model results from uncertainties in
modeling the supergiants, whichmanifest themselves
as the problem of the blue-to-red supergiant ratio
mentioned above.

To estimate how strongly this affects the reconstructed
SFH, we analyzed the sensitivity of the
solution to the choice of stellar evolution models
and metallicity, more specifically, we reconstructed
the SFH using the Padova isochrones with Z =
0.004 and Z = 0.008 and the Geneva isochrones
(Charbonnel et al. 1993) with Z = 0.004. The latter
use the same convection criterion as the Padova
ones. However, as was pointed out by Langer and
Maeder (1995), they give different predictions for
the occurrence frequency of supergiants. A visual
comparison of the two model populations showed
that the locations of the supergiant branches in the 
color-magnitude diagram predicted by the Geneva
isochrones differ significantly from those predicted by
the Padova isochrones. As we see from Fig. 6a, the
solutions obtained with these two models also differ
from one another. Since the observed diagrams are
described by the Padova isochrones much better, the
solution obtained with the latter is probably more realistic
and below we take it as the main one. A similar
situation is also observed for the solutions obtained
with the same (Padova) isochrones, but with different
metallicities -- they are statistically incompatible with
one another, although the differences are appreciably
smaller than than in the case of different stellar
evolution models (Fig. 6b). The general tendency
in the behavior of the solution is retained, because
the main constraints on the SFH are imposed by
the distribution of stars along the main sequence, on
which the stellar evolution is modeled much better
than on the supergiant branches. As a result of the
model inadequacy, the solution also slightly depends
on the choice of the region under consideration in
the color-magnitude diagram. This is illustrated by
Fig. 6b, which shows the solution on a grid with a
more stringent magnitude threshold for supergiants.
As we see from Fig. 6, the solutions differ, but less
than in the previous cases.

\begin{figure*}
\includegraphics[width=0.8\textwidth,clip=true]{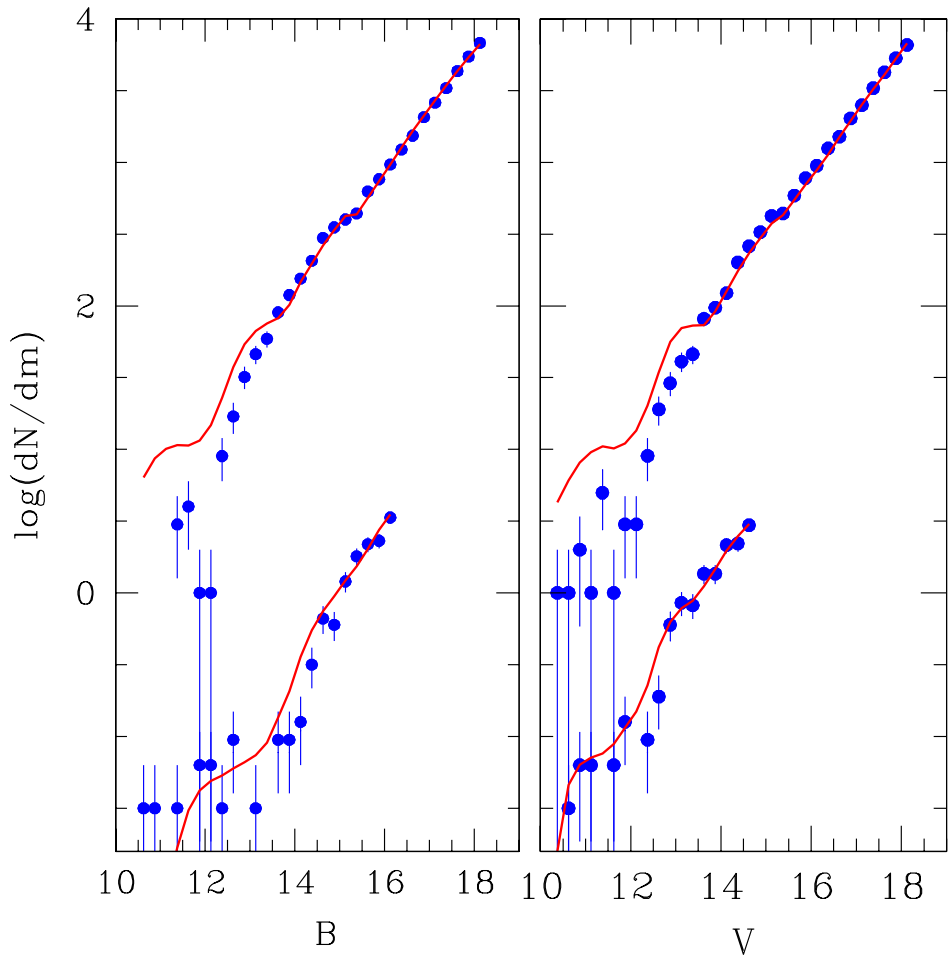}
\caption{Luminosity functions for the stellar population in the SMC for 
the main sequence and blue supergiants (the points with
error bars and the curves in the upper part of the plots) and red supergiants
 (in the lower part of the plots). The points with error
bars represent the observed luminosity function; the curve represents 
their best fits. For clarity, the red supergiant branches
were displaced along the horizontal axis. The excess of the model for
 the main sequence and blue supergiants above the data
in the region of bright stars results from inaccuracy of the currently
 available stellar evolution models for supergiants.
}
\label{fig:lumf}
\end{figure*}

As has already been noted above, we could reconstruct
the SFH using only main-sequence stars,
thereby avoiding the supergiant-related problems.
However, the accuracy of the photometry available at
our disposal is insufficient for the latter to be reliably
separated from the main-sequence stars.

Thus, imperfectness of the models for massive
stars on which the present stellar evolution models
are based limits the reconstruction accuracy of the
recent star formation history. Only the general behavior
of the SFH has a reasonable accuracy, while
the individual features in it should be interpreted with
caution. To minimize the effect of such uncertainties,
below we coarsen the grid in time by combining two
time bins into one. As a result, four of them remain
in the interval log t = 6.6--8.0 instead of eight. As
we see from Fig. 7, although this does not solve
all of the problems considered above, it allows the
uncertainties in the solution related to them to be reduced
appreciably. As will be clear in the subsequent
analysis, a higher time resolution is not required for
the problem under consideration, because the accuracy
of determining the sough-for function $\eta_{HMXB}(t)$
is limited by the Poisson noise associated with the
relatively small number of HMXBs in the SMC.

\begin{figure*}
\begin{centering}\hbox{
\includegraphics[width=0.45\textwidth]{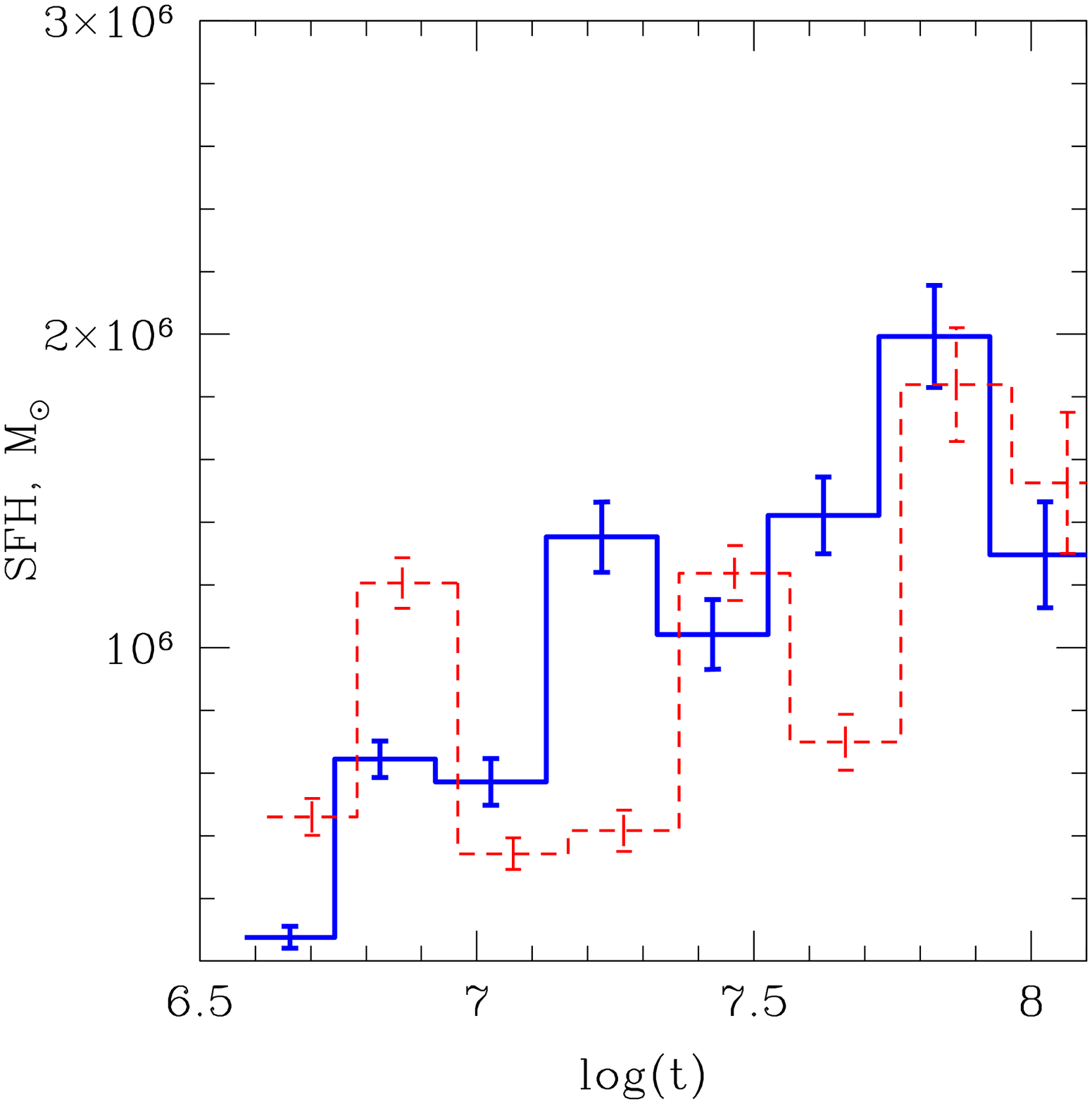}
\includegraphics[width=0.45\textwidth]{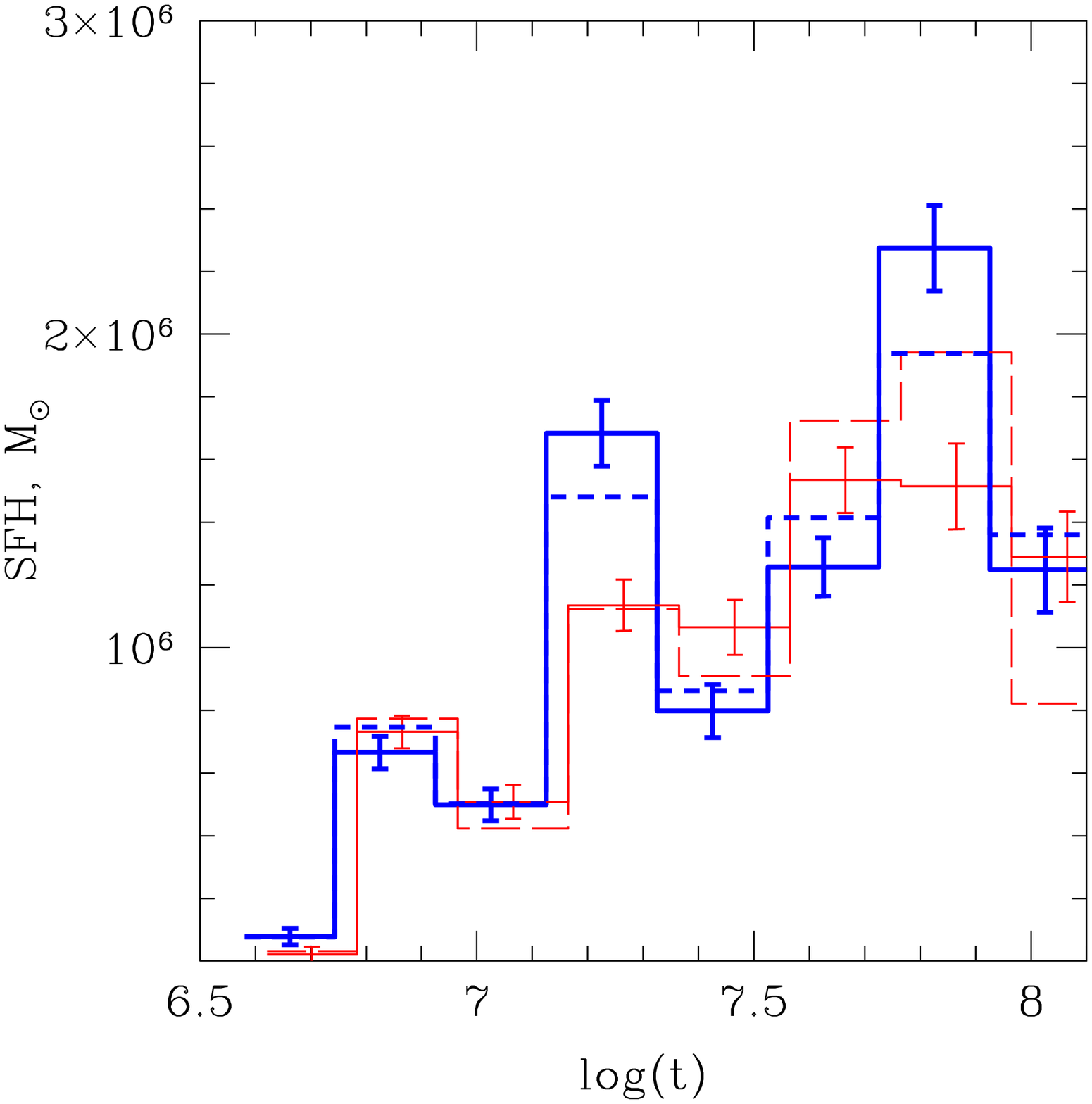}
}\end{centering}
\caption{ (a) Dependence of the solution on the choice of stellar evolution
 models. The solid and dashed histograms correspond
to the solutions obtained from the Padova and Geneva isochrones, 
respectively. Note that the Padova isochrones describe the
distribution of supergiants in the color-magnitude diagram 
much better. (b) Demonstration of the sensitivity of the solution
to metallicity and binning of the color-magnitude diagram. 
The solid thick and thin histograms correspond to the solutions
obtained from the Padova isochrones on themain grid with 
Z = 0.008 and 0.004, respectively. The thick dashed (short dashes)
and thin (long dashes) histograms correspond to the solutions 
obtained with an increased threshold for supergiants and the
same metallicities.
}
\label{fig:sfhgridcmp}
\end{figure*}

\begin{figure*}
\begin{centering}\hbox{
\includegraphics[width=0.45\textwidth]{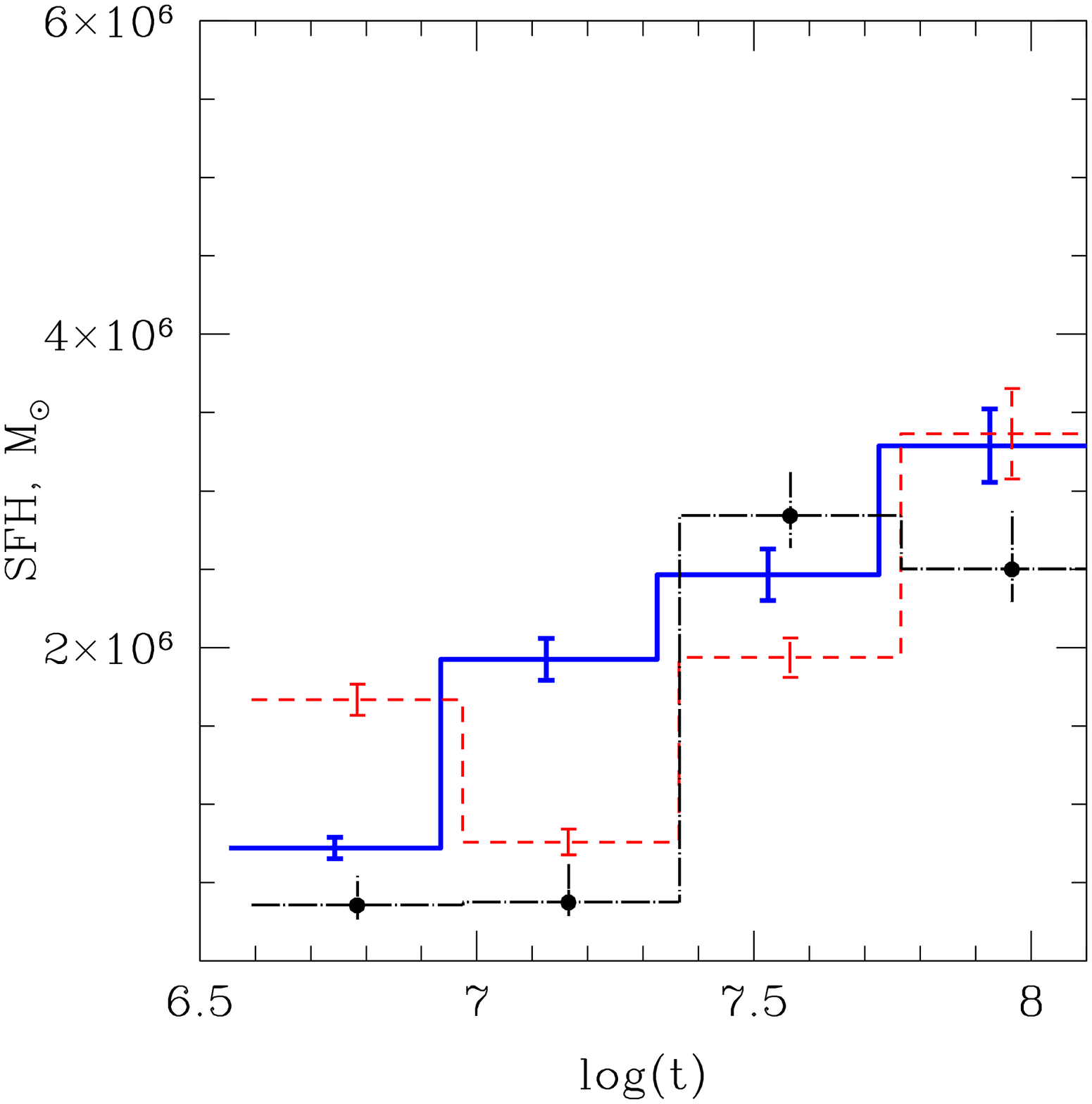}
\includegraphics[width=0.45\textwidth]{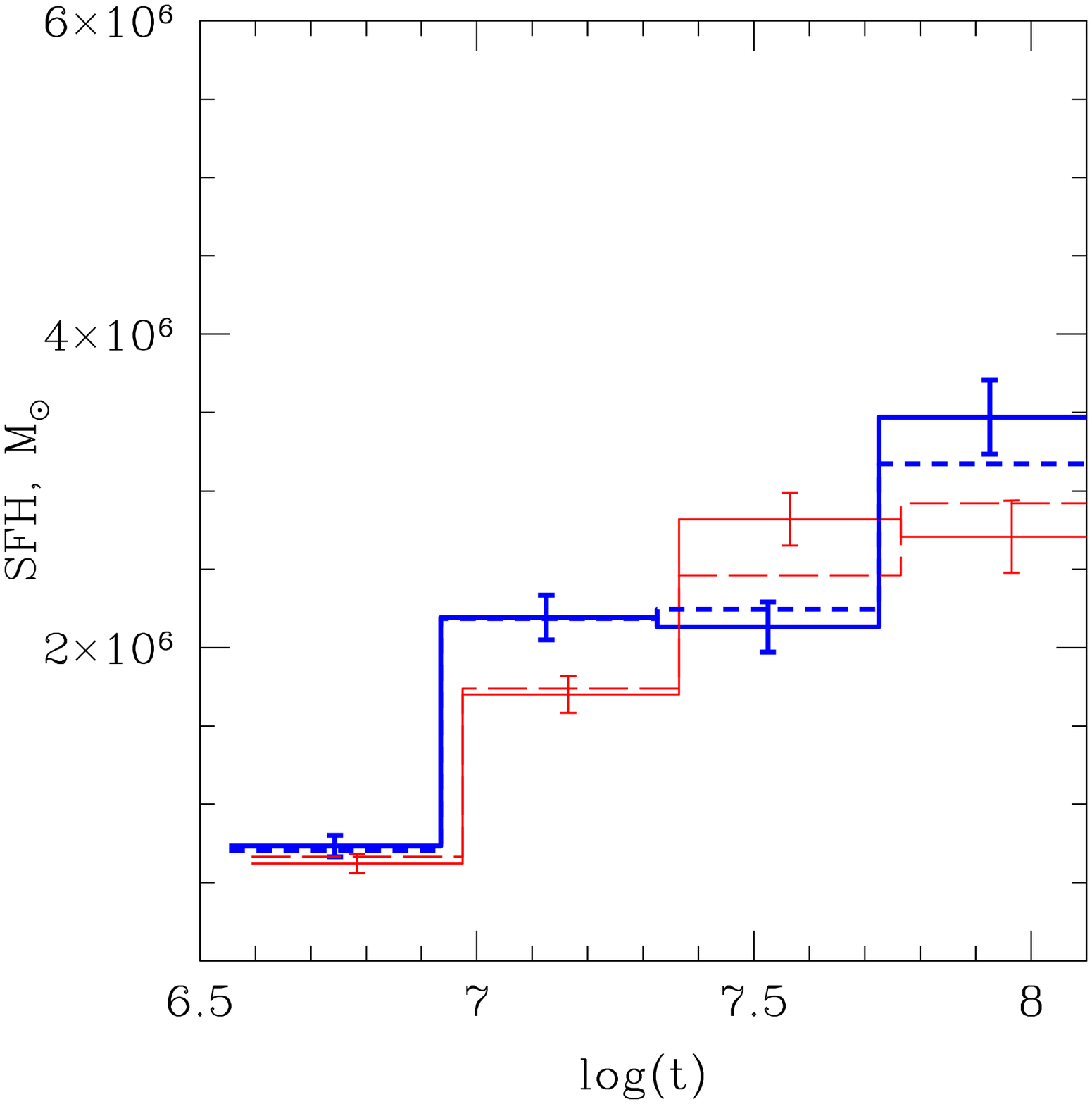}
}\end{centering}
\caption{Same as Fig. 6 constructed in wider time bins. 
Clearly, the problems obvious in Fig. 6 become less prominent
 in the solution with a rougher time resolution. The
 dash-dotted histogram represents the SFH from Harris and Zaritsky (2004).
}
\label{fig:sfhgridcmp2}
\end{figure*}

It is interesting to compare the SFH that we
obtained with that from Harris and Zaritsky (2004),
whose method differs significantly from ours. It
should be kept in mind that Harris and Zaritsky
(2004) investigated the SFH in a wide range of
ages and did not concentrate on the features related
to the reconstruction of recent star formation. The
two SFHs are shown in Fig. 7a. We see that they
are in satisfactory agreement at t$\ga20$~Myr and differ
on shorter time scales. The largest discrepancy is
observed in the second time bin corresponding to
 $\log(t)\approx 7.0-7.3$. For a quantitative comparison of
the accuracies of the two SFHs, let us consider
the number of red supergiants formed in this time
bin predicted by these two dependences. The stars
formed in the bin $\log(t)\approx 7.0-7.3$ that have become
red supergiants by now had initial masses in the
range $\approx12-22M_\odot$. Their current positions in the
color-magnitude diagram are roughly limited by the
intervals of magnitudes V = 12.0--13.5 and colors
B--V=1.4--1.8. The SFH obtained in this paper
predicts 57 stars in these magnitude and color
intervals, while according to Harris and Zaritsky
(2004), their number must be 15. The numbers of
stars are shown for the set of all fields for which the
SFH was obtained in Fig. 7. As would be expected,
the predictions differ by almost a factor of 4. We
emphasize that both solutions are based on the same
stellar evolution models and identical assumptions
about the IMF, the binary fraction, and the stellar
mass distribution in binary systems. These numbers
should be compared with the observed number of red
supergiants, 50. Obviously, the solution obtained in
this paper describes better the population of massive
young stars. Since the mass of the stars formed in this
time bin is low, this difference affects weakly the end
result, as we demonstrate below.

\begin{figure*}
\begin{centering}\hbox{
\includegraphics[width=0.45\textwidth]{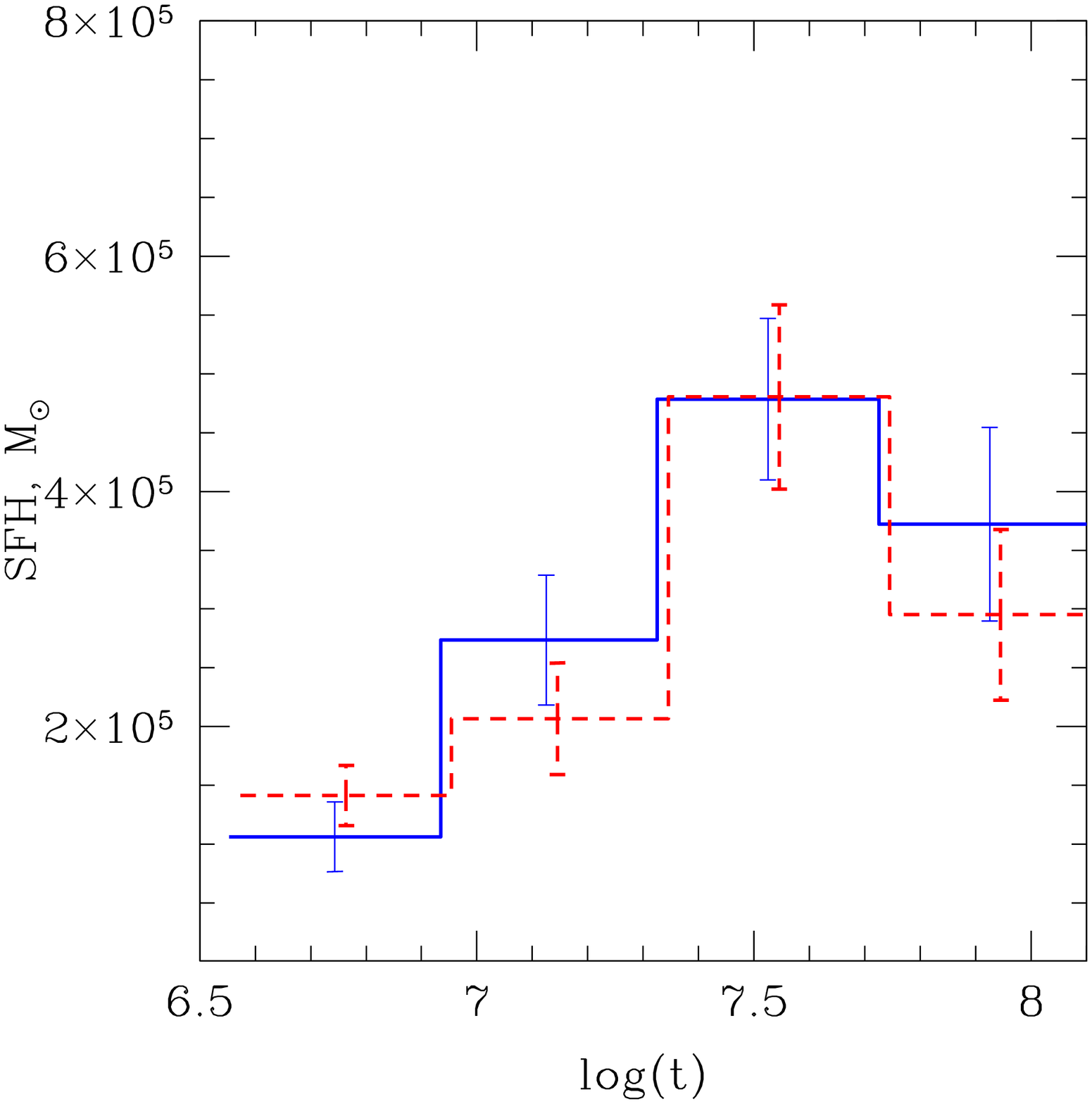}
\includegraphics[width=0.45\textwidth]{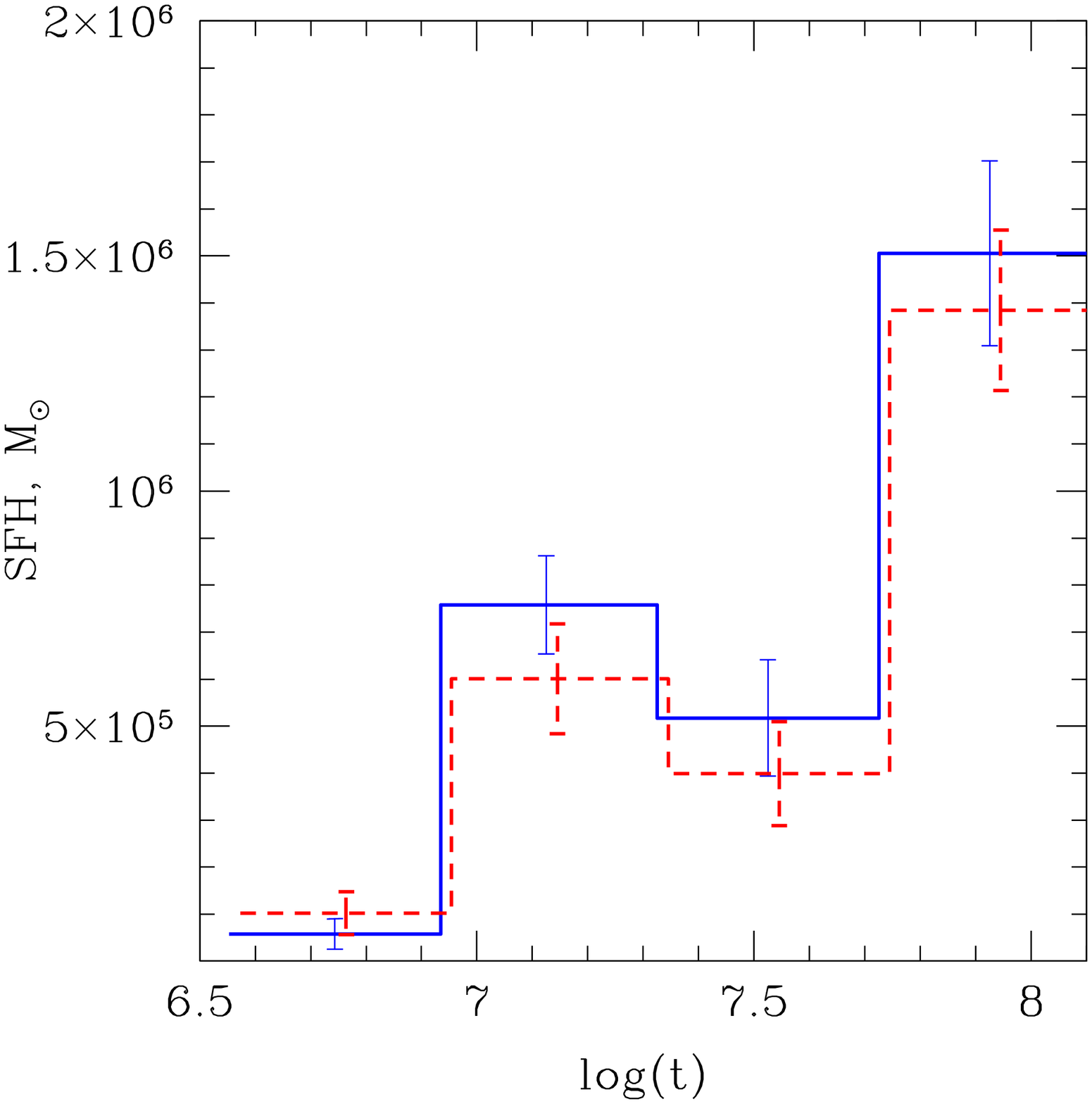}
}\end{centering}
\caption{ Demonstration of the sensitivity of the solution
 to photometric errors. (a) SFH for the actual stellar population in one of
the SMC fields (solid histogram) and the same population 
with photometry distorted by random errors distributed uniformly in
the interval -0.2--0.2 (dashed histogram). 
(b) SFH in one of the fields reconstructed using the MCPS (solid histogram) and
OGLE (dashed histogram) catalogs.
}
\label{fig:sfhscsim}
\end{figure*}

\bigskip
{\it Stability of the solution to photometric errors.}

To verify that the solution is only weakly sensitive
to photometric errors, we performed two tests.
In the first test, we introduced noise into the actual
photometry by shifting the magnitudes by random
values distributed uniformly in the interval from -0.2
to +0.2. Subsequently, we reconstructed the SFH
using the original and distorted photometries. As
is clear from Fig. 8, the solution depends weakly
even on such large errors. As the second test, we
compared the SFHs for the actual stellar population
obtained using two different catalogs, OGLE and
MCPS. Since the OGLE catalog provides photometry
only in the B, V, I bands, we use only the (B-V, V) diagram. 
Obviously, this procedure is equivalent
to reconstructing the SFH for one stellar population
with the errors taken from different distributions. The
derived SFHs are in good agreement with one another
(Fig. 8).

\begin{figure*}
\includegraphics[width=0.8\textwidth,clip=true]{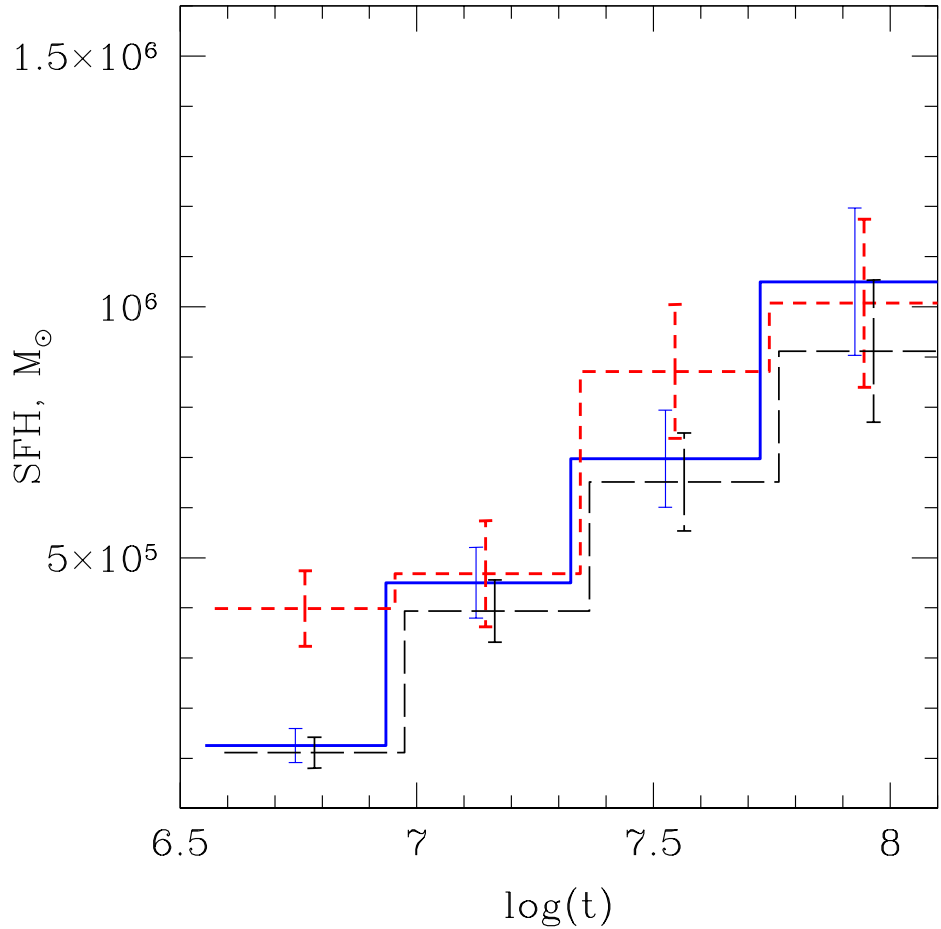}
\caption{Demonstration of the dependence of the solution
on IMF and distribution in binary component mass ratio.
The solid histogramrepresents the solution obtained with
standard parameters; the histograms with long and short
dashes represent the solutions obtained by assuming a
flat distribution in mass ratio and an IMF with a slope
of 2.7, respectively. In the latter case, the normalization
was reduced by a factor of 2.
}
\label{fig:binary}
\end{figure*}

\begin{figure*}
\begin{centering}\hbox{
\includegraphics[width=0.45\textwidth]{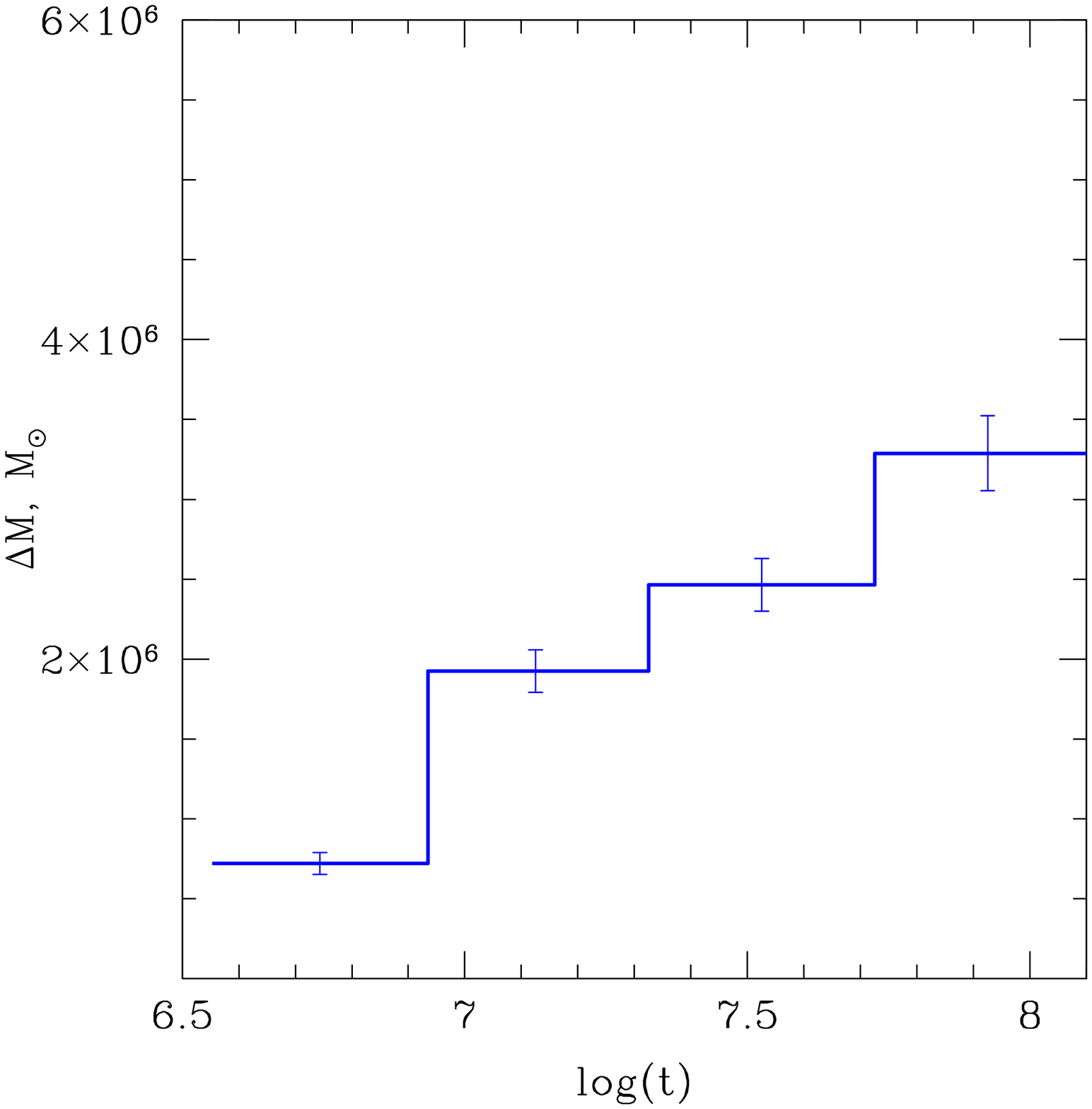}
\includegraphics[width=0.45\textwidth]{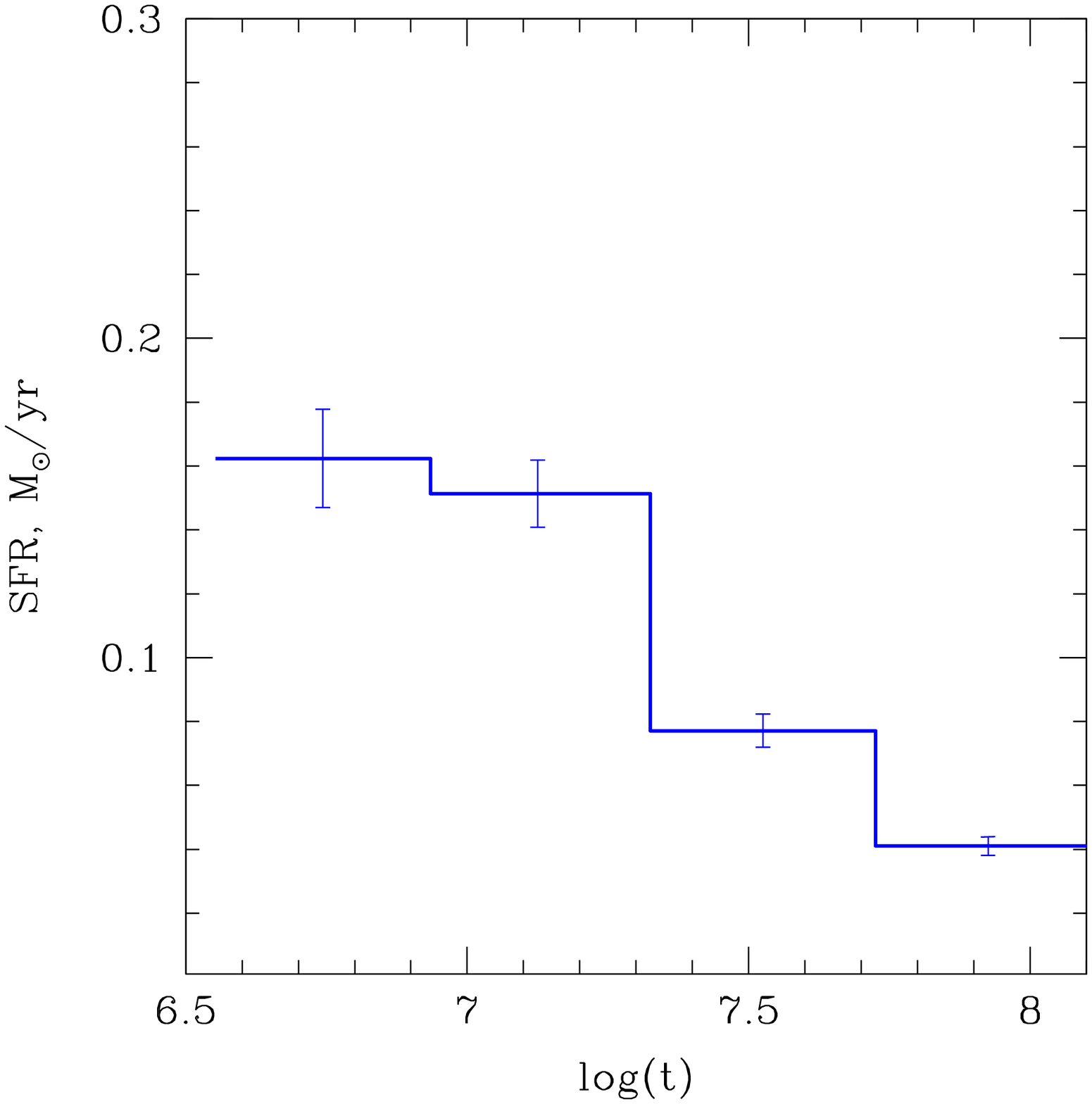}
}\end{centering}
\caption{Combined SFH in the eight SMC regions used to reconstruct the 
dependence N$_{HMXB}$(t). The mass (a) formed in
different time intervals and the SFR (b) are plotted along the vertical 
axes as functions of time.
}
\label{fig:sfhsmc}
\end{figure*}

\bigskip
{\it Dependence of the solution on IMF, binary
fraction, and binary component mass ratio.}

 In constructing the synthetic color\u2013magnitude diagrams,
we assumed a Salpeter IMF. Since we use
only the upper part of the color-magnitude diagram,
only the behavior of the mass function for massive
stars is important to us. Although the stellar mass
distribution in (massive) star clusters is known to
follow the Salpeter mass function up to the highest
masses, the mass function of the field stars may be
steeper (Massey 2003). To check how strongly the
solution depends on the presumed IMF slope, we
reconstructed the SFH in one of the SMC fields by
assuming a steeper slope, $\Gamma=1.7$. The derived SFH
does not differ greatly from the solution obtained
with the standard value of $\Gamma=1.35$, except that its
normalization is a factor of 2 higher (see Fig. 9).
Formally, this difference stems from the fact that we
assume the same IMF slope in the entire mass range
$0.1-100M_\odot$. The end result of our calculations will
be virtually unchanged, since it was normalized to the
total mass of the massive stars $M>8~M_{\odot}$, while the
difference of the normalization is attributable to the
low-mass stars.

In generating the synthetic diagrams, another significant
assumption is made with regard to the fraction
of binary systems and their distribution in mass
ratio. The binary fraction can exceed f$_{binary}$=0.5
adopted here as the standard one, while the distribution
in binary component mass ratio is nearly flat
(see, e.g., Kobulnicky et al. 2006). To analyze the
dependence of the solution on these assumptions, we
reconstructed the SFH for one of the fields by assuming
that the distribution in component mass ratio is
flat and that all stars are in binaries (f$_{binary}$=1). 
In both cases, the solution is found to be close to that
obtained with standard parameters (see Fig. 9; the
solution with a different binary fraction is not provided,
since it is almost identical to the standard one).
However, the conversion coefficient from the number
of stars to the stellar mass depends on these parameters.
Therefore, the normalization of the derived SFH
may differ. Thus, for example, the SFH normalization
for f$_{binary}$=1 increases by a factor of $\approx1.3$.

\subsection{Results: The SFH in the SMC}
\label{sec:sfhresults}

Based on the results of previous sections, we
reconstructed the SFH in the SMC. This was done
separately for each of the regions observed with
XMM-Newton and used in Shtykovskiy and Gilfanov
(2005b) to search for HMXBs. The pointing
at CF Tuc, which is displaced considerably from the
SMC center and contains no HMXBs, constitutes
an exception. The combined SFH for these regions is
shown in Fig. 10.

\begin{figure*}
\begin{centering}\hbox{
\includegraphics[width=0.45\textwidth]{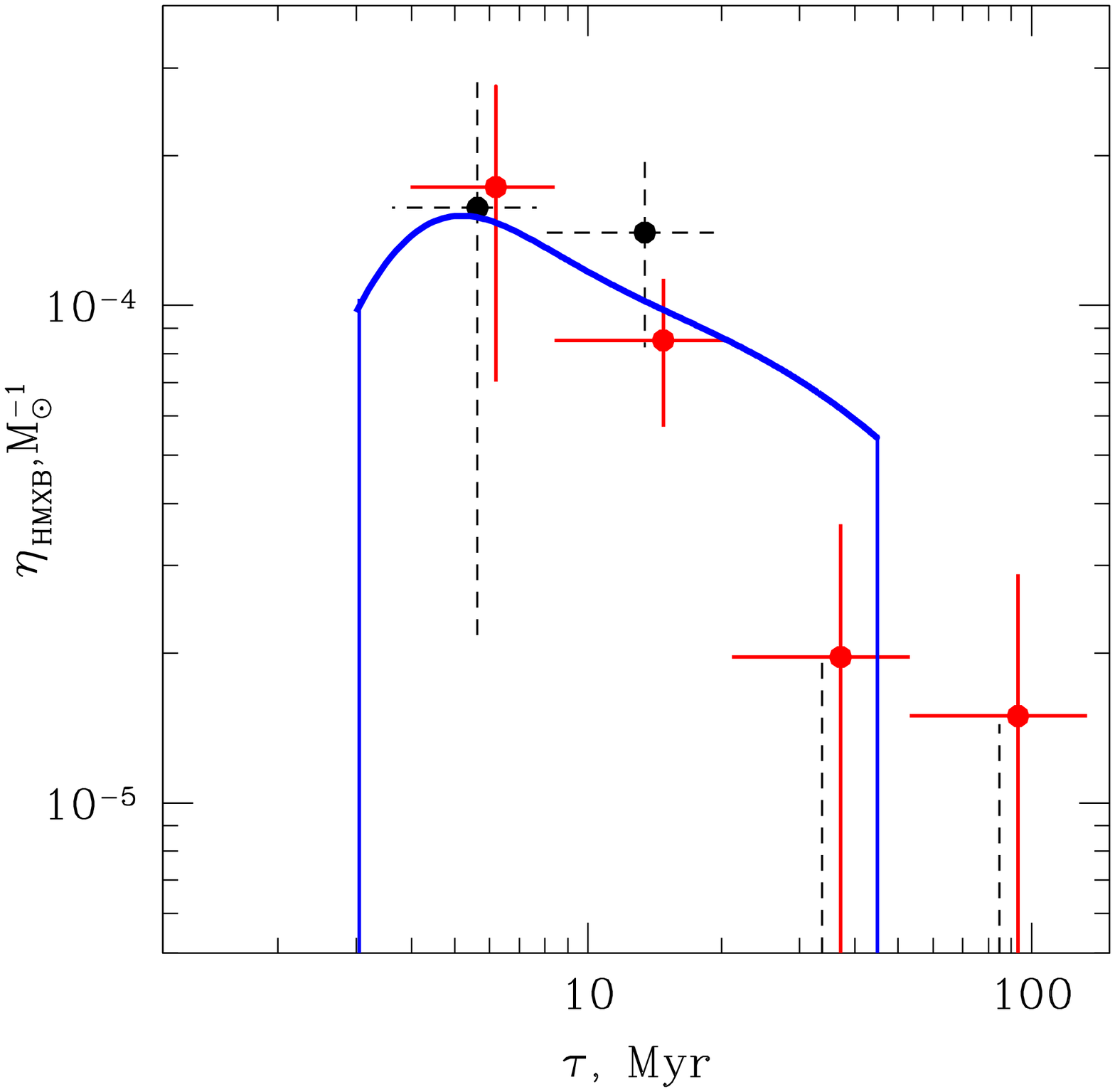}
\includegraphics[width=0.45\textwidth]{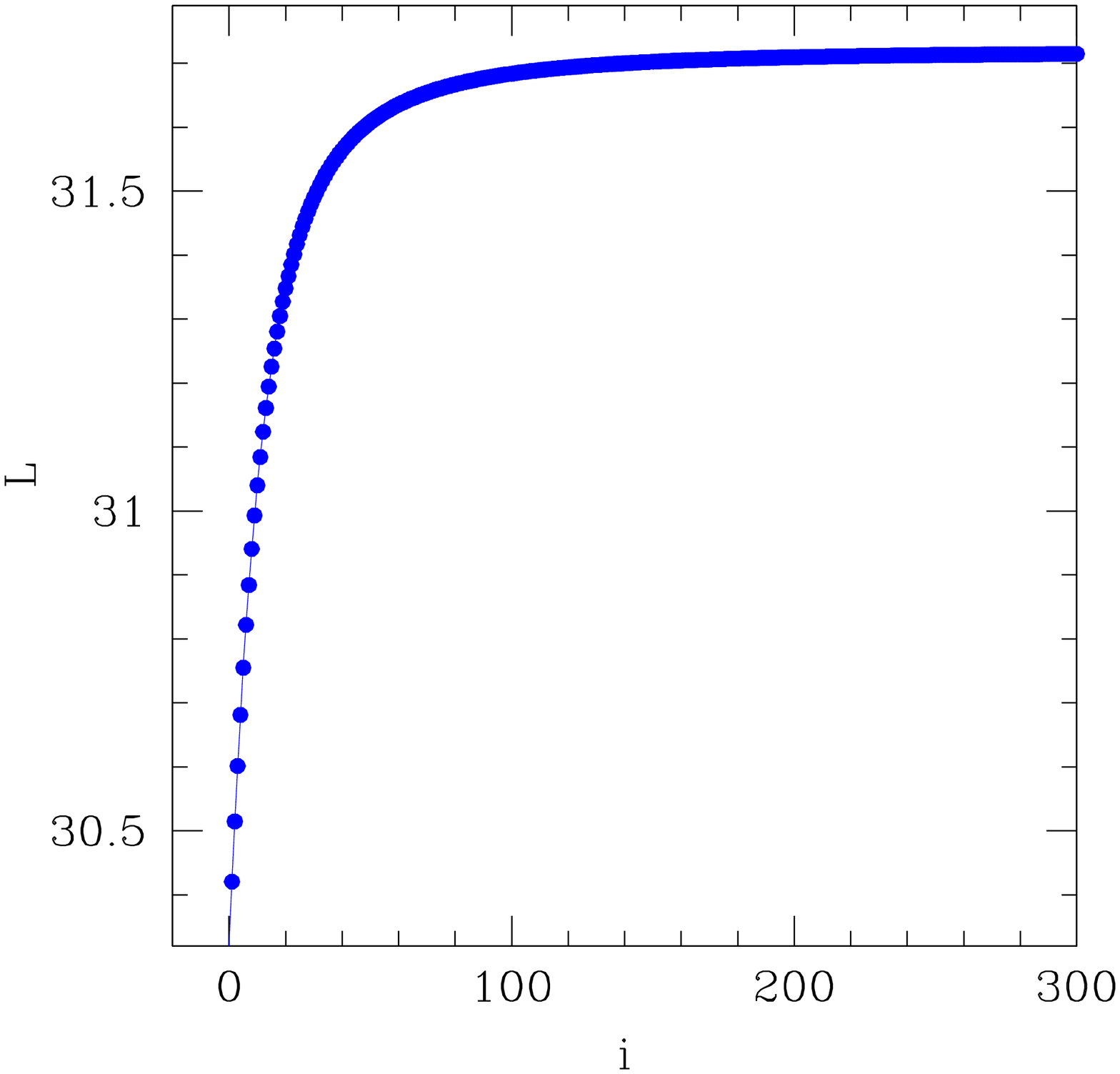}
}\end{centering}
\caption{Demonstration of the reconstruction of themodel dependence of the HMXB
 number on the time elapsed since the star
formation event. (a) The model (solid line) and the solutions obtained by the 
Lucy-Richardson method after 20 (solid crosses)
and 100 (dashed crosses) iterations. (b) Likelihood function of the solution 
versus number of iterations.
}
\label{fig:etahmxbsim}
\end{figure*}

\begin{figure*}
\begin{centering}\hbox{
\includegraphics[width=0.45\textwidth]{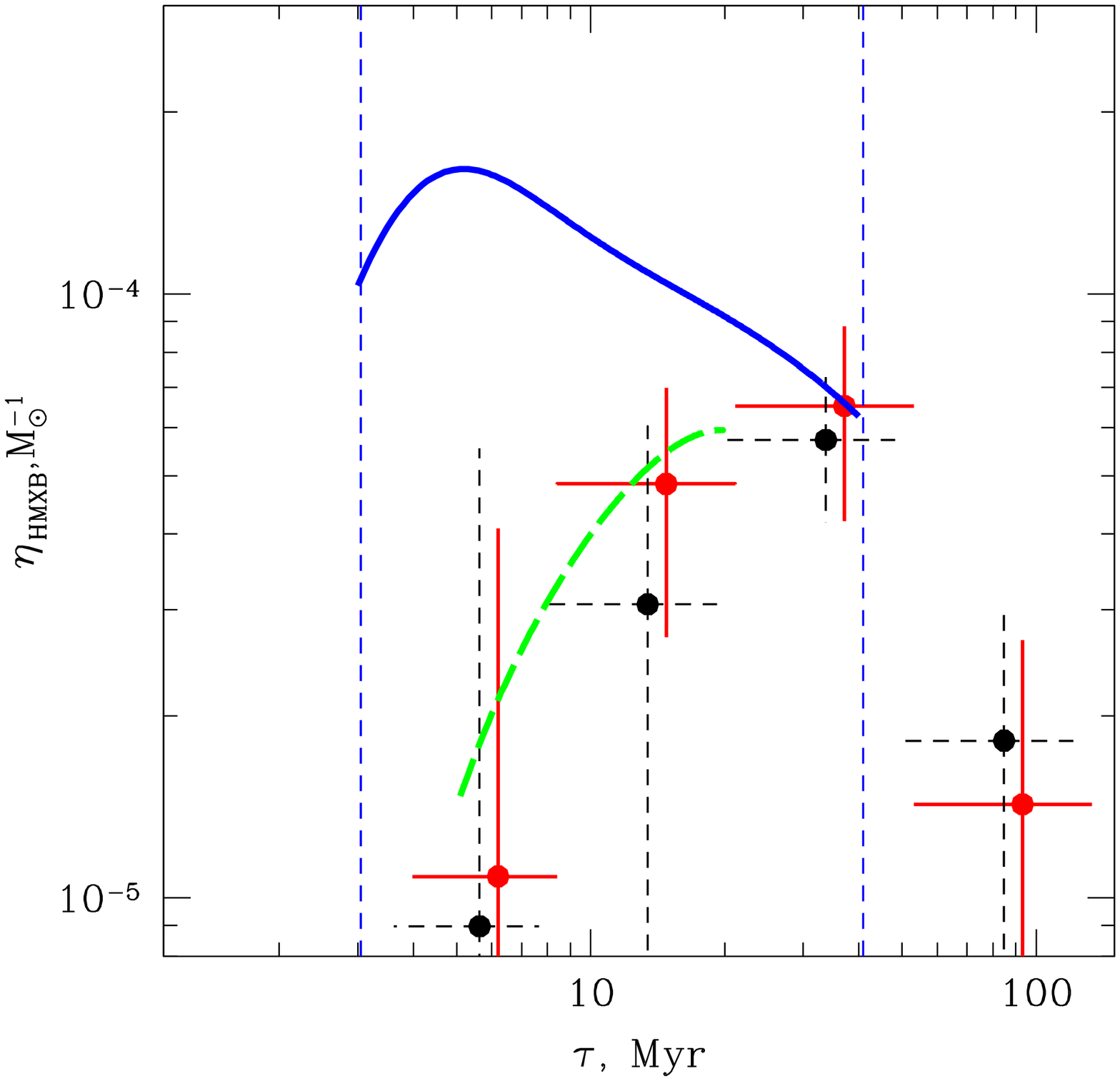}
\includegraphics[width=0.45\textwidth]{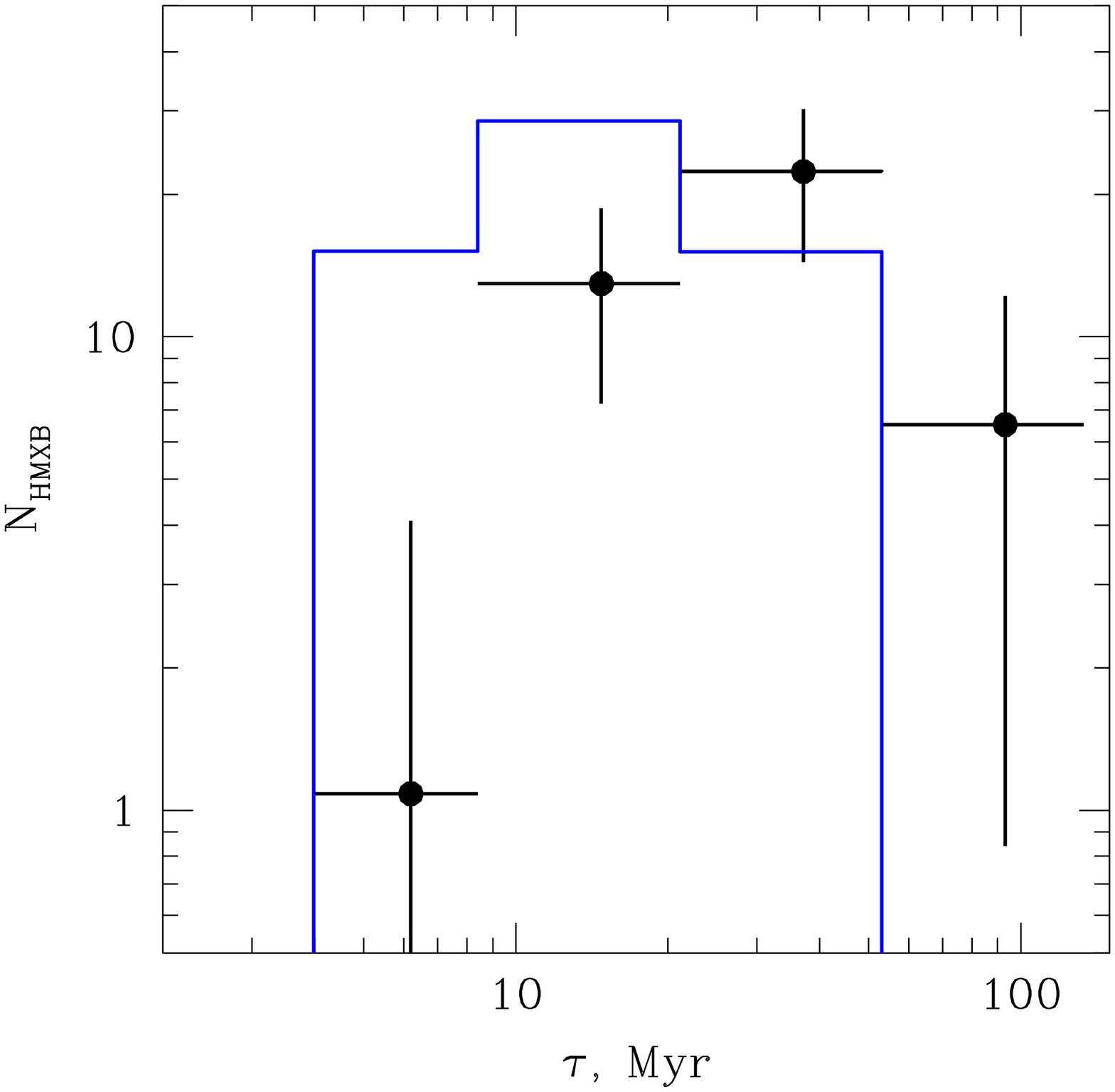}
}\end{centering}
\caption{(a) Dependence of the HMXB number on the time elapsed since the 
star formation event. The solid and dashed
crosses were obtained using the SFH reconstructed in this paper and the SFH from
 Harris and Zaritsky (2004). The solid
curve represents the model based on the supernova rate. The two vertical dashed
 lines reflect the formation times of the first
black hole and the last neutron star calculated in terms of the standard theory of
 evolution of a single star. The dashed curve
represents the theoretical dependence of the number of Be/X systems with neutron stars 
from Popov et al. (1998). (b) The
age distribution of HMXBs in the SMC. The points with error bars were obtained by 
multiplying the data of the observational
curve (a) by the mass formed in the corresponding time bins. The histogram reflects 
the predictions of the model based on the
supernova rate. Obviously, the observed number of the youngest systems is 
appreciably lower than the predicted one.
}
\label{fig:etahmxb}
\end{figure*}

\section{EVOLUTION OF THE HMXB POPULATION
AFTER THE STAR FORMATION EVENT} 
\label{sec:etahmxbevol}

Having the spatially resolved star formation history
SFR(t,X), let us turn to the solution of Eq. (4).
To construct the function $N_{HMXB}(t,X)$, we used our
catalog of HMXBs in the SMC (Shtykovskiy and
Gilfanov 2005b), from which we selected HMXBs
brighter than $10^{34}$~erg/s. This threshold corresponds
to the detection of $\approx$75\% of the sources (see
Shtykovskiy and Gilfanov 2005b). To take into account
the incompleteness of the catalog, we will divide
our solution $\eta_{HMXB}(t)$ by 0.75. The spatial variable
X in Eq. (4) is basically the index numbering the
XMM-Newton fields of view -- Eq. (4) is written for
each pointing.After discretization, we obtained a system
of eight linear equation for four unknowns. The
number of unknowns is determined by the number of
time bins in the interval log t = 6.6--8.0.
As above, we used the iterative Lucy--Richardson
method to obtain a regularized solution. To find the
stopping criterion, we solved the problem based on
the model function $\eta(t)$. Using the SFHs SFR(t,X)
and the function $\eta(t)$ based on the SN II rate, we
calculated the expected numbers of HMXBs in eight
spatial regions in the SMC and their Poisson realizations.
The total number of model sources is close
to the actual number of HMXBs in the SMC. The
reconstructed dependence $\eta(t)$ is shown in Fig. 11
for two stopping criteria; one is close to the plateau
in the likelihood function (100 iterations) and the
other is long before it (20 iterations). As we see from
the figure, the latter is in better agreement with the
model and, in addition, has smaller errors. This means
that the problem is ill-posed and its solution requires
regularization.

\section{RESULTS AND DISCUSSION}
\label{sec:etahmxbevolfin}

The derived dependence of the HMXB number
on the time elapsed since the star formation event,
$\eta_{HMXB}(t)$, is shown in Fig. 12a. The uncertainty of
the solution was calculated in the same way as above
in the Subsection ``Uncertainty of the Solution''. To
take into account the incompleteness of the catalog of
HMXBs in the SMC, we multiplied the normalization
of the solution obtained by the factor 1.3. The theoretical
curve in Fig. 12a corresponds to the model based
on the SN II rate and normalized using the N$_{HMXB}$--SFR
 calibration (Grimm et al. 2003), as described in
the Section ``Evolution of the HMXB Population after
the Star Formation Event''. The solutions shown in
Fig. 12a were obtained both for the SFH determined
in this paper and for the SFH from Harris and Zaritsky
(2004). We see that the solutions are compatible,
despite a certain difference between the two SFHs
in the lower time bins (Fig. 7) -- as was mentioned
above, the accuracy of the solution is limited by the
Poisson noise due to the relatively small number of
HMXBs in the SMC.

As is clear from Fig. 12a, the HMXB formation
efficiency does not exceed the prediction based on the
mean N$_{HMXB}$--SFR relation for the local Universe.
The abundance of HMXBs in the SMC is the result
of a specific form of the recent SFH in this galaxy,
namely, its high rate $\sim50$~Myr ago.

The specific form of $\eta_{HMXB}(t)$   differs significantly
from the behavior of the SN II rate: the HMXB number
reaches its maximum 20--50 Myr after the star
formation event, i.e., on time scales of the order of or
longer than the explosion time of the last supernova
with the formation of a neutron star. Note also the
paucity of the youngest systems compared to the
model predictions. Obviously, most of the young systems
correspond to HMXBs with black holes, since
they are the first to be formed after the star formation
event. This shortage is not unexpected from an observational
point of view, since most of the HMXBs
in the SMC are known to be pulsars with Be companions.
However, it is of great interest from the
standpoint of the theory of formation and evolution of
binary systems. Obviously, this behavior is related to
the evolution of a companion star whose lifetime can
reach $\sim 60$~Myr for a single $6M_{\odot}$ star (when the evolution
effects in the binary system are disregarded).
Another important factor is the evolution of the neutron
star spin period (Illarionov and Sunyaev 1975).
Population synthesis models are an adequate tool for
studying these effects. As an example, Fig. 12a shows
the time dependence of the number of Be/X systems
with neutron stars derived by Popov et al. (1998)
based on calculations using the ``Scenario Machine''.
The systems of other classes (e.g., neutron stars with
supergiants) are much less numerous, given the luminosity
threshold of  10$^{33}$~erg/s chosen by the
authors. Therefore, we provide no curves for them.
To be able to compare the absolute number of Xray
sources with the results of our observations, we
renormalized the theoretical dependence to the number
of systems brighter than $10^{34}$~erg/s. For this
purpose, we used the luminosity function for HMXBs
in the SMC obtained previously (Shtykovskiy and
Gilfanov 2005b). Note that its slope in the range of
low luminosities is slightly smaller than the standard
value of 0.6 (Grimm et al. 2003). As we see from
Fig. 12a, there is good agreement with the observations
both in the shape of the dependence and in
its normalization in the time interval 5--20 Myr in
which the models by Popov et al. (1998) are valid.
Note that Popov et al. (1998) performed their calculations
by assuming a solar heavy-element abundance,
while the details of the population of X-ray sources
depend on metallicity (Dray 2006). Obviously, our
experimental dependence can be used to test and
``calibrate'' the population synthesis models and to
clarify the various aspects of the evolution of binary
systems.

Figure 12b shows the age distribution of HMXBs
in the SMC, which is the product of the reconstructed
dependence $\eta_{HMXB}(t)$ by the mass of the
stars formed in the corresponding time bins. As is
clear from Fig. 12b, the HMXB population in the
SMC is rather old, $\tau\approx20-50$~Myr. Dray (2006)
also reached a similar conclusion by analyzing the
observed distributions of HMXB periods and luminosities
and by comparing them with the results of
population synthesis models. She also suggested the
existence of a relatively recent intense star formation
event in the SMC.

When the results shown in Fig. 12 are interpreted,
it should be kept in mind that we used X-ray
sources with luminosities L$_X\geq10^{34}$erg/s, i.e.,
faint sources dominate in our sample, to reconstruct
the time dependence of the HMXB number. It would
be interesting to look at the behavior of the function
 $\eta_{HMXB}(t)$ for bright sources, e.g.,  L$_X\ga10^{37}$erg/s.
Indeed, although the luminosity of a specific binary
depends on the size of its orbit, one may expect its
mean X-ray luminosity to rise with increasing mass
of the companion star. Bright X-ray binaries will
then be, on average, younger than faint ones due
to the shorter lifetime of more massive stars. This
conclusion is also supported by the observational fact
that brighter sources in star-forming galaxies are,
on average, closer to young star clusters (see, e.g.,
Kaaret al. 2004). Therefore, the time dependence of
the number of bright sources will differ from that
shown in Fig. 12. However, such a study cannot be
performed for the SMC because of its insufficiently
high SFR and, accordingly, small number of bright
sources. Note also that the N$_{HMXB}$--SFR and L$_X$--SFR
 relations from Grimm et al. (2003) are based
on Chandra observations of bright HMXBs in other
galaxies. Therefore, one might expect these relations
to break down for a lower luminosity threshold. This
effect will be unimportant for the L$_X$--SFR relation,
since the total X-ray luminosity of the HMXB population
is determined mainly by bright sources in view
of the shape of their luminosity function. However,
the total number of sources is determined by the more
numerous faint sources. Therefore, one might expect
noticeable deviations from a linear relation in the
N$_{HMXB}$--SFR relation when the luminosity threshold
is lowered.

\section{CONCLUSIONS}
\label{sec:discussion}

We considered the relation between the HMXB
population and the SFH of the host galaxy. The number
of HMXBs can be represented as a convolution
(Eq. (4)) of the star formation history SFR(t) with
the function $\eta_{HMXB}(t)$ describing the dependence of
the HMXB number on the time elapsed since the star
formation event. Thus, the evolution of the HMXB
population after the star formation event can be reconstructed
by analyzing the distribution of HMXBs
in stellar complexes with different SFHs.

Using archival optical observations, we reconstructed
the spatially resolved SFH in the SMC over
the past $\sim$100~Myr (Fig. 10). For this purpose, the
observed color-magnitude diagrams of the stellar
population were approximated by linear combinations
of model isochrones. We analyzed the stability and
errors of this method for reconstructing the recent
SFH and showed that its accuracy is limited by the
uncertainties in the currently available models for the
evolution of massive stars. However, the systematic
error introduced by this factor may be ignored, since
the main source of uncertainty in the solution is the
Poisson noise due to the relatively small number
of HMXBs in the part of the SMC investigated by
XMM-Newton.

Using the derived SFHs and the spatial distribution
of HMXBs in the SMC from Shtykovskiy
and Gilfanov (2005b), we reconstructed the function
$\eta_{HMXB}(t)$ that describes the dependence of the
HMXB number on the time elapsed since the star
formation event (Fig. 12). We compared the derived
dependence with the behavior of the SN II rate. The
HMXB number reaches its maximum $\sim$20--50~Myr
after the star formation event, which is comparable
to or exceeds the lifetime of a  $8M_\odot$ star. This is
much later than the maximum of the SN II rate. In
addition, note the shortage of the youngest systems.
Observationally, this manifests itself in the absence
(or an extremely small number) of HMXBs with black
holes in the SMC. This behavior is related to the
evolution of the companion star and the neutron star
spin period and is consistent with the population synthesis
model calculations (Popov et al. 1998). When
these results are interpreted, it should be kept in mind
that the function $\eta_{HMXB}(t)$ depends on the luminosity
threshold used to select the X-ray sources. In
our analysis, we used a sample with a low luminosity
threshold, L$_{min}\sim 10^{34}$~erg/s. In such a sample,
low-luminosity sources, mostly Be/X systems,
mainly contribute to the number of sources, while the
relative contribution from systems with black holes
and/or O/B supergiants, which must constitute the
majority of sources in the lower time bin in Fig. 12, is
small. Therefore, the time dependence of the number
of bright sources (e.g.,  $>10^{37}$~erg/s) will differ from
that shown in Fig. 12.

The HMXB formation efficiency in the SMC does
not exceed the prediction of the N$_{HMXB}$--SFR calibration
(Grimm et al. 2003). Their abnormal abundance
compared to the predictions based on the emission
in standard SFR indicators, such as the H$_{\alpha}$ line,
can result from a peculiarity of the SFH in the SMC.

\section*{ACKNOWLEDGMENTS}
P.E. Shtykovskiy thanks the European Association
for Research in Astronomy (EARA;
MESTCT-2004-504604) for support by a Marie
Curie fellowship and the Max-Planck-Institut fur Astrophysik,
where much of this work was performed, for
hospitality. This work was also supported by grant
no. NSh-1100.2006.2 from the President of Russia
and the Russian Academy of Sciences (``Origin and
Evolution of Stars and Galaxies'' Program). We
thank the referees for valuable remarks that helped
to noticeably improve the presentation of our results
in the paper.

\pagebreak


\begin{thebibliography}{}

\bibitem[\protect\citeauthoryear{Aparicio et al.}{1997}]{aparicio97}
Aparicio, A., Gallart, C., Bertelli, G., Astron. J. 114, 680 (1997).

\bibitem[\protect\citeauthoryear{Belczynski et al.}{2005}]{belczynski05}
Belczynski, K., Kalogera, V., Rasio, F.A. et al. (astro-ph/0511811) (2005).


\bibitem[\protect\citeauthoryear{Westerlund}{1997}]{mc_book97}  
Westerlund, B., ``The Magellanic Clouds'', Cambridge, New York: 
Cambridge Univ.Press, 1997.


\bibitem[\protect\citeauthoryear{Gallart et al.}{2005}]{gallart05}
Gallart, C., Zoccali, M., Aparicio, A., 
Ann. Rev. Astron. Astrophys. 43, 387 (2005).


\bibitem[\protect\citeauthoryear{Dolphin}{1997}]{dolphin97}
Dolphin, A., New Astron. 2, 397 (1997).

\bibitem[\protect\citeauthoryear{Dolphin}{2000}]{dolphin00}
Dolphin, A., MNRAS 313, 281 (2000).

\bibitem[\protect\citeauthoryear{Dolphin}{2002}]{dolphin02}
Dolphin, A., MNRAS 332, 91 (2002).

\bibitem[\protect\citeauthoryear{Dohm-Palmer et al.}{1997}]{dohm-palmer97}
Dohm-Palmer, R.C., Skillman, E.D., Saha, A. et al., 
Astron. J. 114, 2527 (1997).

\bibitem[\protect\citeauthoryear{Dray}{2006}]{dray06}
Dray, L. M., MNRAS 370, 2079 (2006).


\bibitem[\protect\citeauthoryear{Grimm et al.}{2003}]{grimm03}
Grimm, H.-J., Gilfanov, M., Sunyaev, R., MNRAS 339, 793 (2003).

\bibitem[\protect\citeauthoryear{Girardi et al.}{2002}]{girardi02}
Girardi, L., Bertelli, G., Bressan, A. et al.,  Astron. Astrophys. 391, 195 (2002).

\bibitem[\protect\citeauthoryear{Zaritsky et al.}{2002}]{zaritsky02}  
Zaritsky, D., Harris, J., Thompson, I.B. et al., Astron. J. 123, 855 (2002).

\bibitem[\protect\citeauthoryear{Zaritsky et al.}{2004}]{zaritsky04}  
Zaritsky, D., Harris, J., Thompson, I.B. et al., 
Astron. J. 128, 1606 (2004).

\bibitem[\protect\citeauthoryear{Illarionov \& Sunyaev}{1975}]{illarionov75}  
Illarionov, A. F., Sunyaev, R. A., Astron. Astrophys.
39, 185 (1975).


\bibitem[\protect\citeauthoryear{Kaaret et al.}{2004}]{kaaret04}
Kaaret, P., Alonso-Herrero, A., Gallagher, J.S. et al., 
MNRAS 348, 28 (2004).


\bibitem[\protect\citeauthoryear{Kobulnicky et al.}{2006}]{kobulnicky06}  
Kobulnicky, H., Fryer, C., Kiminki, D., astro-ph/0605069.


\bibitem[\protect\citeauthoryear{Langer \& Maeder}{1995}]{langer95}
Langer, N., Maeder, A., Astron. Astrophys 295, 685 (1995).

\bibitem[\protect\citeauthoryear{Lucy}{1974}]{lucy74}
Lucy, L.B., Astron. J. 79, 745 (1974).

\bibitem[\protect\citeauthoryear{Lucy}{1994}]{lucy94}
Lucy, L.B., Astron. Astrophys. 289, 983 (1994). 

\bibitem[\protect\citeauthoryear{Massey}{2002}]{massey02}
Massey, P., Astrophys. J. Suppl. Ser.  141, 81 (2002).

\bibitem[\protect\citeauthoryear{Massey}{2003}]{massey03}
Massey, P., Ann. Rev. Astron. Astrophys. 41, 15 (2003).


\bibitem[\protect\citeauthoryear{Maeder et al.}{1999}]{maeder99}
Maeder, A., Grebel, E. K., Mermilliod, J.-C., Astron. Astrophys., 346, 459 (1999)


\bibitem[\protect\citeauthoryear{Popov \& Prokhorov}{2004}]{popov04}
Popov, S.B., Prokhorov, M.E.,  
Helmholtz International Summer School and Workshop 
on Hot points in Astrophysics and Cosmology (astro-ph/0411792) (2004).

\bibitem[\protect\citeauthoryear{Popov et al.}{1998}]{popov98}
Popov, S.B., Lipunov, V.M., Prokhorov, M.E. et al., 
Astron. Rep. 42, 1, 29 (1998).

\bibitem[\protect\citeauthoryear{Pagel \& Tautvaisiene}{1998}]{pagel98}
Pagel, B.E.J., Tautvaisiene, G., MNRAS 299, 535 (1998).


\bibitem[\protect\citeauthoryear{Russell \& Dopita}{1992}]{russel92}
Russell, S.C., Dopita, M.A., Astrophys. J. 384, 508 (1992).


\bibitem[\protect\citeauthoryear{Udalski et al.}{1998}]{udalski98}  
Udalski, A., Szymanski, M., Kubiak, M. et al., Acta Astron. 48, 147 (1998).

\bibitem[\protect\citeauthoryear{Harris \& Zaritsky}{2004}]{harris04}
Harris, J., Zaritsky, D., Astron. J. 127, 1531 (2004).

\bibitem[\protect\citeauthoryear{Holtzman et al.}{1999}]{holtzman99}
Holtzman, J.A., Gallagher, J.S. III, Cole, A.A. et al., 
Astron. J. 118, 2262 (1999).

\bibitem[\protect\citeauthoryear{Charbonnel et al.}{1993}]{charbonnel93}
Charbonnel, C., Meynet, G., Maeder, A. et al., 
Astron. Astrophys. Suppl. Ser. 101, 415 (1993).

\bibitem[\protect\citeauthoryear{Shtykovskiy \& Gilfanov}{2005а}]{shtykovskiy05a}
Shtykovskiy, P., Gilfanov, M., 
Astron. Astrophys. 431, 597 (2005).

\bibitem[\protect\citeauthoryear{Shtykovskiy \& Gilfanov}{2005б}]{shtykovskiy05b}
Shtykovskiy, P., Gilfanov, M., MNRAS 362, 879 (2005).


\bibitem[\protect\citeauthoryear{Schaller et al.}{1992}]{schaller92}
Schaller, G., Schaerer, D., Meynet, G. et al., Astron. Astrophys. Suppl. Ser. 96, 269 (1992).

\end{thebibliography}
\end{document}